\documentclass[twocolumn,prd,superscriptaddress,nofootinbib]{revtex4-1}

\usepackage{graphicx}
\usepackage{amsmath,bm}
\usepackage[usenames,dvipsnames]{xcolor}
\usepackage[normalem]{ulem}
\usepackage{url}
\usepackage{multirow}
\usepackage[colorlinks  = true,
            linkcolor   = NavyBlue,
            urlcolor    = NavyBlue,
            citecolor   = NavyBlue,
            anchorcolor = NavyBlue]{hyperref}

\def \us {ODDK22}

\newcommand{\sigmaInv}{\sigma_{\rm inv}}

\newcommand{\etal}{{\it et al.}}

\def \apj  {ApJ}

\def \nat {Nature}

\def \etal {et~al.~}

\begin{document}

\title{New determination of the production cross section for $\gamma$ rays in the Galaxy}

\author{Luca Orusa}
\email{luca.orusa@edu.unito.it}
\affiliation{Department of Physics, University of Torino, via P. Giuria, 1, 10125 Torino, Italy}
\affiliation{Istituto Nazionale di Fisica Nucleare, via P. Giuria, 1, 10125 Torino, Italy}

\author{Mattia Di Mauro}
\email{dimauro.mattia@gmail.com}
\affiliation{Istituto Nazionale di Fisica Nucleare, via P. Giuria, 1, 10125 Torino, Italy}

\author{Fiorenza Donato}
\email{donato@to.infn.it}
\affiliation{Department of Physics, University of Torino, via P. Giuria, 1, 10125 Torino, Italy}
\affiliation{Istituto Nazionale di Fisica Nucleare, via P. Giuria, 1, 10125 Torino, Italy}

\author{Michael Korsmeier}
\email{michael.korsmeier@fysik.su.se}
\affiliation{The Oskar Klein Centre for Cosmoparticle Physics, Department of Physics, Stockholm University, Alba Nova, 10691 Stockholm, Sweden}

\begin{abstract}
\noindent
The flux of $\gamma$ rays is measured with unprecedented accuracy by the \emph{Fermi} Large Area Telescope from 100 MeV to almost 1 TeV. In the future, the Cherenkov Telescope Array  will have the capability to measure photons up to 100 TeV. To accurately interpret this data, precise predictions of the production processes, specifically the cross section for the production of photons from the interaction of cosmic-ray protons and helium with atoms of the ISM, are necessary.
In this study, we determine new analytical functions describing the Lorentz-invariant cross section for $\gamma$-ray production in hadronic collisions. We utilize the limited total cross section data for $\pi^0$ production channels and supplement this information by drawing on our previous analyses of charged pion production to infer missing details. In this context, we highlight the need for new data on $\pi^0$ production. Our predictions include the cross sections for all production channels that contribute down to the 0.5\% level of the final cross section, namely $\eta$, $K^+$, $K^-$, $K^0_S$, and $K^0_L$ mesons as well as $\Lambda$, $\Sigma$, and $\Xi$ baryons. 
We determine the total differential cross section $d\sigma(p+p\rightarrow \gamma+X)/dE_{\gamma}$ from 10 MeV to 100 TeV with an uncertainty of $10\%$ below 10 GeV of $\gamma$-ray energies, increasing to 20\% at the TeV energies. 
We provide numerical tables and a script for the community to access our energy-differential cross sections, which are provided for incident proton (nuclei) energies from 0.1 to $10^7$ GeV (GeV/n).
\end{abstract}

\maketitle

\section{Introduction}

Gamma rays ($\gamma$ rays) represent the most energetic photons produced in the Universe and only the most powerful astrophysical processes can generate them. In the last 15 years, the Large Area Telescope (LAT) on board NASA's Fermi Gamma-ray Space Telescope ({\it Fermi}) \cite{Fermi-LAT:2009ihh} has revolutionized the field of $\gamma$-ray astronomy providing data with unprecedented precision. \emph{Fermi}-LAT is a satellite-based experiment integrating a silicon tracker with an electric calorimeter. It has detected about  6500 $\gamma$-ray sources over the full sky and in an energy range from 100 MeV up to about 1 TeV \cite{Fermi-LAT:2019yla,Ballet:2020hze,Fermi-LAT:2022byn}. {\it Fermi}-LAT data have been used by several groups to study non-thermal radiation processes that produce high-energy photons in the Universe.
Ground-based experiments take advantage of their larger collective area and extend the energy range up to PeV scales. They all use the Earth's atmosphere as a calorimeter and, exploiting different techniques, they are able to disentangle air showers produced by cosmic rays and very-high-energy $\gamma$ rays. Photons are identified using either Cerenkov telescopes, like Magic \cite{MAGIC:2014zas}, H.E.S.S. \cite{DeNaurois:2020bac}, and the forthcoming Cerenkov Telescope Array (CTA) \cite{CTAConsortium:2013ofs}, or water Cerenkov detectors like HAWC~\cite{HAWC:2020hrt} or LHAASO~\cite{LHAASO:2019qtb,2021Natur.594...33C}. 

Modern detectors measure the arrival direction of $\gamma$ rays with up to $0.1^{\circ}$ precision. Since they travel on straight lines, $\gamma$ rays can be used to do astronomy. The most numerous $\gamma$-ray sources in our Galaxy are pulsars and supernova remnants. In the extragalactic sky, several thousand blazars have been identified as point-like sources, while others like mAGN \cite{DiMauro:2013xta} or SFG \cite{Tamborra:2014xia,Roth:2021lvk} are mostly too weak to be identified individually and only a few ones have been detected. However they are numerous enough to contribute significantly to the extragalactic $\gamma$-ray background \cite{Fornasa:2015qua,DiMauro:2015tfa}. Furthermore, many transient sources like GRBs have been observed \cite{Ajello:2019zki}.

Most of the $\gamma$ rays detected by {\it Fermi}-LAT are produced by the Galactic insterstellar emission, which is generated by the interaction of charged cosmic rays (CRs) with the atoms of the interstellar gas or the low-energy photons of the interstellar radiation fields \cite{Fermi-LAT:2012edv}. The dominant processes, especially at low latitudes, are the hadronic interactions of CR nuclei with the gas of the Galactic disk \cite{Fermi-LAT:2016zaq,
Porter:2017vaa,Kissmann:2017ghg,Johannesson:2018bit,Tibaldo:2021viq,Dundovic:2021ryb,Widmark:2022qgx}.  Typically it is called the ``$\pi^0$-component'' because most, although not all, of the $\gamma$ rays in hadronic interactions, originate from the decay of $\pi^0\rightarrow \gamma\gamma$. In addition to the hadronic processes, $\gamma$ rays are also produced by the bremsstrahlung, i.e. when CR electrons and positrons scatter on the Galactic gas atoms, or when low-energy photons are up-scattered to $\gamma$ rays through so-called inverse Compton scattering \cite{Strong:1998fr}. 

The hadronic $\gamma$-ray diffuse emission is determined by the inelastic scattering of CR nuclei (mostly proton and helium) against interstellar medium (ISM) atoms at rest in the Galactic disc. The rate of interactions depends on the CR fluxes, the density of the ISM, and the inelastic production cross section $\sigma(p+p\rightarrow \gamma +X)$ (and similarly for a nuclear components in CRs and in the ISM). The local CR fluxes are measured with high accuracy by AMS-02 \cite{AMS:2021nhj}, and the density of ISM in our local environment, i.e. within a couple of kpc, is known with good precision~\cite{Widmark:2022qgx}. Further away from the solar position, the situation becomes more complicate because CR fluxes are subject to extrapolations from local measurements and, therefore, very model dependent. Also the gas distribution far from local Galaxy is more difficult to determine. There, the gas density is typically obtained by combining the Doppler-shift information in 21 cm maps with a modeling of the gas rotation around the Galaxy~\cite{Pohl:2007dz,Mertsch:2022oee}.

There are several indications for our incomplete knowledge of the non-local Galaxy coming from the modeling of the $\gamma$-ray sky. The typical approach of the modeling is to construct templates in concentric rings around the Galactic plane. Those templates are then fitted to $\gamma$-ray data using free normalizations for each energy bin \cite{Fermi-LAT:2014ryh}. Usually, either the CR model or the gas model are altered significantly by the template fit, often making them incompatible with (local) expectations. One example is the observation of a hardening in the $\gamma$-ray spectrum towards the Galactic center \cite{Fermi-LAT:2016zaq,Yang:2016jda,Pothast:2018bvh}. Improving the modeling of the $\gamma$-ray sky is a central topic in current research. In any case, a key ingredient to properly predict the hadronic $\gamma$-ray diffuse emission is the inclusive $\gamma$-ray production cross section $\sigma(p+p\rightarrow \gamma +X)$. Any uncertainty in these cross sections comparable or greater than the $Fermi$-LAT statistical errors undermine the study of the Galactic interstellar emission  of the observed $\gamma$-ray sky. 

In this work, we investigate the cross section of hadronic interactions to produce $\gamma$ rays, with the aim to estimate its correct dependence on kinematic variables and robustly size the error bars inherent the modeling. 
Since data are very limited for these cross sections, the standard approach is to determine them employing Monte Carlo event generators \cite{Kamae:2006bf,Kachelriess:2019ifk,Bhatt_2020,Mazziotta_2016}. The most commonly used cross section parametrization relies on a customized implementation of Pythia 6 by Kamae \etal\ \cite{Kamae:2006bf}. Another, more recent result has been provided by AAfrag~\cite{Kachelriess:2019ifk} based on the QGSJET-II-04m event generator, which is specifically tuned to high energies. 

There can be significant deviations between Monte Carlo simulations and experimental data as demonstrated by \cite{Kachelriess:2015wpa,Kachelriess:2019taq} for the production cross sections of $\bar{p}$ and in Ref.~\cite{Orusa:2022pvp} (hereafter \us) for a few channels for the production of $e^{\pm}$. Moreover, as shown in Ref~\cite{Koldobskiy:2021nld} the differences in the production cross sections of $\gamma$ rays obtained with different Monte Carlo generators can be even larger than $30\%$. This demonstrates the necessity of improving the model of these cross sections.

We present in this paper a new and more precise model relying mostly on an analytic prescription. The main production mechanism of $\gamma$ rays is the decay of $\pi^0$ mesons. However, data for the $\pi^0$ production are extremely scarce. There are measurements of the multiplicity, but data on the Lorentz-invariant differential cross section are either not given or affected by large systematics or do not cover the kinematic region relevant for Astroparticle physics.
Therefore, we decide to fit the multiplicity of $\pi^0$ and extrapolate the kinematics from the production cross section of $\pi^{+}$ and $\pi^-$ by taking a combination of the parametrizations obtained in \us. As a consequence, we face larger systematic uncertainties which are intrinsically difficult to quantify. We, therefore, encourage further experimental efforts to measure neutral pion production in hadronic collisions. Then, we carefully model also the production cross sections of $\eta$ and $K$ mesons and $\Lambda$ baryons, which contribute significantly to the $\gamma$-ray cross sections through direct production of photons or with the decay into $\pi^0$ mesons.
This strategy closely follows the one from \us\ where we derived cross sections for the secondary production of CR electrons and positrons. Our approach is similar to the formalism used in \cite{Moskalenko:1997gh} but with much better cross section data for inelastic proton-proton scattering and pion production.

We note that the most obvious application of the cross section from this work is the computation of the diffuse $\gamma$-ray background. Thus, in the following, we will use it as a benchmark to exemplary show the implications of our work. However, we anticipate that the cross section is actually important also in a much larger context. It is required as input for modeling most of the point sources mentioned above. 
For example, a fraction of $\gamma$ rays from blazars is believed to be produced by inelastic hadronic interactions \cite{Matthews:2020lig}.  Moreover, whenever individual $\gamma$-ray (point) sources are studied, the $\pi^0$-component forms an important background~\cite{Fermi-LAT:2019yla}. A prominent example is the Galactic Center where a significant excess of $\gamma$ rays has been observed and discussed controversially in the last decade in the $15\times15$~deg region of interest around the Galactic Center \cite{Hooper:2010mq,Fermi-LAT:2017opo}. This excess is suppressed by about 2.5 orders of magnitude compared to the $\pi^0$-component. Thus, an accurate prediction of the diffuse background and the  $\pi^0$-component is crucial. In this sense, almost every $\gamma$-ray analysis relies either directly or indirectly on the cross section we investigate in the following.

The remainder of the article is structured as follows. In Sec. \ref{sec:emissivity} we outline the theoretical elements to derive the observed flux of $\gamma$ rays from the cross section for the production of photons from hadronic processes. Sec. \ref{sec:pi0} is devoted to the analytical modeling of the $\pi^0$ production cross section, which is the main contributor to hadronic $\gamma$ rays. In Sec. \ref{sec:other_channels} we estimate the contribution from other production channels and from scatterings involving nuclei. We presented our results in Sec. \ref{sec:results}, before drawing conclusions in Sec. \ref{sec:conclusions}. 

\begin{figure*}[t]
  \centering {
    \includegraphics[width=0.80\textwidth,trim={0 5cm 0 0},clip]{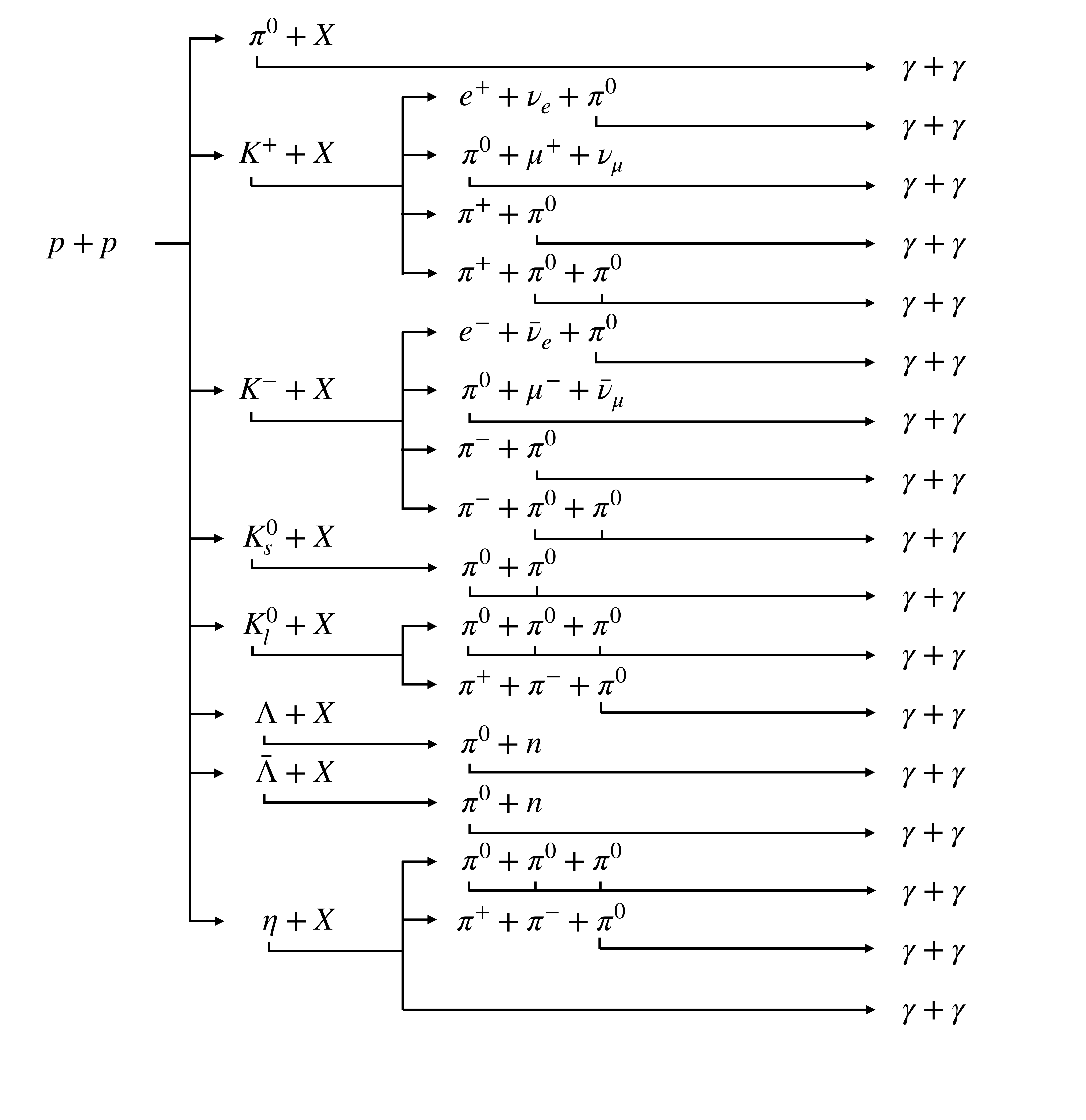}
  }
  \caption{
             This diagram shows the $\gamma$-ray production channels from $p+p$ collisions considered in our analysis.
              We report here only the channels that produce at least 0.5\% of the total yield (see the main text for further details).
             \label{fig:gamma_production_channels}
  }
\end{figure*}

\section{From cross sections to the $\gamma$-ray emissivity}
\label{sec:emissivity}

We briefly summarize in this section the calculation of the hadronic $\gamma$-ray flux. As discussed above, the hadronic component is a very important -- often the dominant -- contribution of the $\gamma$-ray flux. 
The flux detected at Earth $\phi_{\gamma}$ is given by the line of sight (l.o.s.) integral of the $\gamma$-ray emissivity $\epsilon^{i,j}$ and the sum over all interactions of CR species $i$ and ISM components $j$:
\begin{eqnarray}
    \label{eq:gamma_ray-flux}
    \frac{d^2 \phi_\gamma}{d \Omega d E_\gamma} (E_\gamma, l, b) 
    = 
    \sum\limits_{ij}  \int\limits_{\text{l.o.s.}} \!\!\!\!\! \;d \ell \;\; \epsilon^{ij}\Big(\Vec{x}(\ell, l,b), E_\gamma\Big) \, .
\end{eqnarray}
Here $\ell$ is the distance along the l.o.s. while $l$ and $b$ are longitude and latitude. The $\gamma$-ray flux is differential in $\gamma$-ray energy, $E_\gamma$, and solid angle, $\Omega$.
The emissivity at each location in the Galaxy $\Vec{x}$ is the convolution of the CR flux $\phi_i$ and the ISM density $n_{\rm{ISM},j}$ with the energy-differential cross section for $\gamma$-ray production $d\sigma_{ij}/dE_{\gamma}$ for the reaction $i+j\rightarrow \gamma + X$:
\begin{equation}    
    \label{eq:source_term}
    \epsilon^{ij}( \Vec{x}, E_\gamma) = n_{\mathrm{ISM},j}(\Vec{x})
    \int dT_i \, \phi_i(\Vec{x},T_i)\frac{d\sigma_{ij}}{d E_{\gamma}}(T_i,E_{\gamma}) \,.
\end{equation}
We note that, in general, the emissivity depends on the position in the Galaxy since both the ISM gas density and the CR flux are a function of the position.

The vast majority of $\gamma$-ray photons are not directly produced in the proton-proton (or nuclei) collisions but rather by the decay of intermediate mesons and hadrons. In Fig.~\ref{fig:gamma_production_channels}, we show a sketch of all the production channels considered in this analysis. 

The dominant channel is the production of neutral pions, $\pi^0$, and their subsequent decay into two photons. This channel is discussed in detail in Sec.~\ref{sec:pi0}, while we address the contributions from all other channels listed in Fig.~ \ref{fig:gamma_production_channels} in Sec. \ref{sec:other_channels}. Channels that contribute less than $0.5\%$ to the total $\gamma$ production are not shown and neglected in this work. Some of these contributions are not well known and it is difficult to quantify the exact global amount, but we expect that they add up to about 1\% which is well within our uncertainty estimate.

The $\gamma$-ray production cross section is derived from the $\pi^0$ production cross section as follows:
\begin{equation}    
    \label{eq:convolution}
    \frac{d\sigma_{ij}}{d E_{\gamma}}(T_i,E_{\gamma})= 
     \int  d T_{\pi^0} \, \frac{d\sigma_{ij}}{dT_{\pi^0}}(T_i,T_{\pi^0})
     \; P(T_{\pi^{0}}, E_{\gamma}) \, ,
\end{equation}
where $T_{\pi^0}$ is the kinetic energy of the neutral pion decaying into a photon with energy $E_{\gamma}$. The probability density function $P(T_{\pi^0}, E_{\gamma})$ of the process can be computed analytically. 

The fully differential production cross section is defined in the following Lorentz invariant form:
\begin{equation}
    \sigma^{(ij)}_{ {\rm inv}}= E_{\pi^{0}} \frac{d^3 \sigma_{ij}}{dp_{\pi^{0}}^3}\,.
    \label{eq:invariant}
\end{equation}
Here $E_{\pi^{0}}$ is the total $\pi^{0}$ energy and $p_{\pi^{0}}$ its momentum. The fully differential cross section is a function of three kinematic variables, for example, the center of mass energy $\sqrt{s}$, the transverse momentum of the pion $p_T$, and the radial scaling $x_R$. The latter is defined as the pion energy divided by the maximal pion energy in the center of mass frame (CM, labeled by a $*$), $x_R = E_{\pi^{0}}^\ast/E_{\pi^{0}}^{\max\ast}$.

The energy-differential cross section  in Eq.~\eqref{eq:convolution} is obtained by first transforming the kinetic variables from CM into the fix-target (LAB) frame, and then by integrating over the solid angle $\Omega$: 
\begin{eqnarray}
    \frac{d\sigma_{ij}}{d T_{\pi^{0}}}(T_i,T_{\pi^{0}}) &=& p_{\pi^{0}}\int d \Omega \; \sigma_{ {\rm inv}}^{(ij)}(T_i,T_{\pi^{0}},\theta) \\
    &=& 2\pi p_{\pi^{0}} \int^{+1}_{-1} d(\cos{\theta}) \; \sigma_{ {\rm inv}}^{(ij)}(T_i,T_{\pi^{0}},\theta)\,, \nonumber
    \label{eq:solid_int}
\end{eqnarray}
where $\theta$ is the angle between the incident projectile and the produced $\pi^{0}$ in the LAB frame.
We now discuss the $\gamma$-ray production cross sections from $\pi^0$ , benefiting from the results obtained in \us\ for the $\pi^\pm$ production cross sections.

\section{$\gamma$ rays from $p+p \rightarrow \pi^0 + X$ collisions}
\label{sec:pi0}

Given the relevance of the $\pi^0$ channel for the $\gamma$-ray production, it would be important to have precise data on a wide coverage of the kinematic phase space for the reaction $p+p \rightarrow \pi^0 +X$. 
Unfortunately, the available data are either not given for the double differential cross sections or affected by large systematics or do not cover the kinematic region relevant for Astroparticle physics. Instead, for the process $p+p \rightarrow \pi^\pm +X$ data for $\sigma_{\rm{inv}}$ has been collected by various experiments and large portions of the kinetic parameter space, as for example by NA49~\cite{2005_NA49} or NA61~\cite{Aduszkiewicz:2017sei} . 
Therefore, we decide to model $\sigma_{\rm{inv}}$ for the production of $\pi^0$ using the results of $\pi^\pm$ cross sections that we derived in \us. More specifically, we assume that the shape of the $\pi^0$ cross section lies in between the $\pi^+$ and $\pi^-$ shape. Then, we will use the difference between the $\pi^+$ and the $\pi^-$ cross section to bracket the uncertainty as further detailed below.

\subsection{Model for the invariant production cross section}

We assume that $\sigmaInv$ depends on kinematic variables by a relation between the shapes of the production cross sections of $\pi^{+}$ and $\pi^{-}$ as derived in \us, to which we refer for more details.
Thus, for $p+p$ scattering we define $\sigmaInv$ as:
\begin{eqnarray}
       \label{eq:main_equation}
       \sigma_{\rm inv} &=& \sigma_0 (s) \, c_{20} \, \Big[G_{\pi^+}(s, p_T, x_R) \\ \nonumber 
       && \qquad\qquad\qquad +\, G_{\pi^-}(s, p_T, x_R)\Big] \, A(s),
\end{eqnarray}
where $\sigma_0(s)$ is the total inelastic $p+p$ cross section, the functions $G_{\pi^+}$ ($G_{\pi^-}$) represent the kinematic shapes of the invariant $\pi^+$ ($\pi^-$) cross section, and $c_{20}$ is an overall factor that adjusts the total normalization of the cross section. 
The functions $G_{\pi^\pm}(s,p_T,x_R)$ are taken from \us. Their exact definition is:
\begin{eqnarray}
   \label{eq:auxiliary}
   G_{\pi^\pm}(s,p_T,x_R) &=&  c_{1,\pi^\pm} \,   \Big[F_{p,\pi^\pm}(s, p_T, x_R) \\ \nonumber 
   &&\qquad \qquad \qquad + F_{r,\pi^\pm}(p_T, x_R)\Big]
\end{eqnarray}
with $c_1$, $F_p$, and $F_r$ specified in Eqs.~(7) through (9) of \us. We note that the dependence of $F_{p,\pi^\pm}(s, p_T, x_R) $ on $\sqrt{s}$ is extremely mild. The parameters $c_1$ to $c_{19}$ in the definitions of  $G_{\pi^\pm}$ are fixed to the values stated in \us\ (Tab. 2).
Finally, the factor $A(s)$ allows adjusting the cross section to the measured $\pi^0$ multiplicities at different incident energies:
\begin{eqnarray}
  \label{eq:As}
  A(s) &=&   \left(1+\left(\sqrt{s/c_{21}}\right)^{c_{22}-c_{23}}\right)
  \left(1+\left(\sqrt{s/c_{24}}\right)^{c_{23}-c_{25}}\right)\nonumber \\ 
  &&\left(1+\left(\sqrt{s/c_{26}}\right)^{c_{25}-c_{27}}\right) 
  \left(\sqrt{s}\right)^{c_{27}} \biggl/ A(\sqrt{s_0}), 
\end{eqnarray}
where $\sqrt{s_0}$ is fixed to $17.27$~GeV, while the parameters from $c_{20}$ to $c_{27}$ are derived in this work.

\subsection{Fit to total cross section at different $\sqrt{s}$}
\label{sec:piplus_differentCME}

The kinematic shape of the invariant $\pi^0$ production cross section with respect to $p_T$ and $x_R$ has been fixed in the previous section. Here we focus on the scaling of the cross section at different $\sqrt{s}$. Our parametrization introduces the dependence on $\sqrt{s}$ through the function $A(s)$ that acts as an overall renormalization. In this section, we proceed with the determination of the parameters from $c_{20}$ to $c_{27}$. 
To obtain a complete dependence from $\sqrt{s}$ we use the collection of total $\pi^0$ cross section measurements provided in Ref.~\cite{dermer1986binary} (in the following also called Dermer86) and initially compiled in \cite{Stecker73}.

\begin{figure*}[t]
    \includegraphics[width=0.49\textwidth]{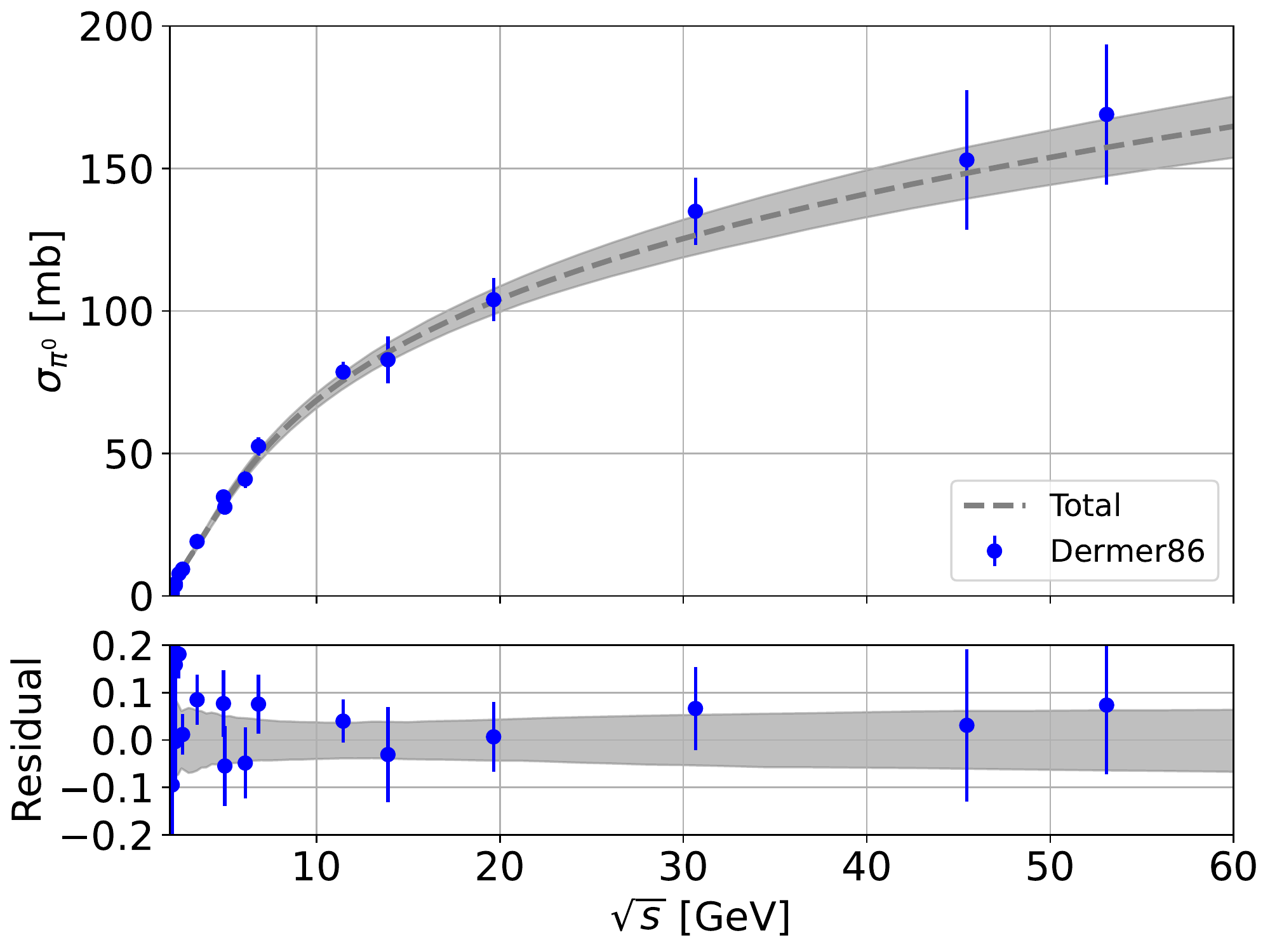}
    \includegraphics[width=0.49\textwidth]{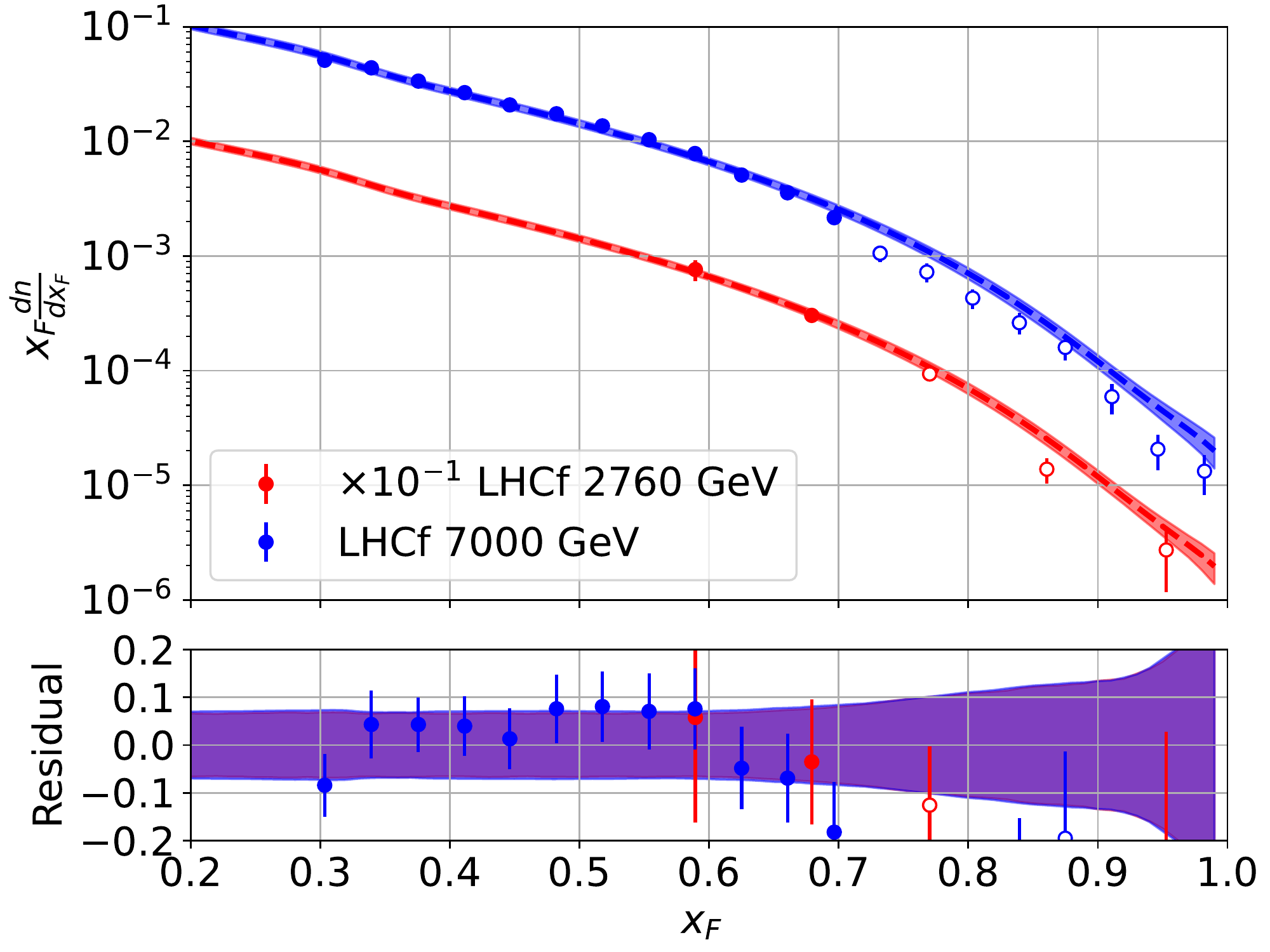}
    \caption{Total cross section (left panel) and $x_F dn/dx_F$ (right panel) of $\pi^0$ production in $p+p$ collisions measured at different $\sqrt{s}$. The solid lines represent the best-fit parametrization and the shaded bands show the uncertainty of our fit at the $1\sigma$ level. The bottom panels shows the residuals defined as (data-model)/model.
    } 
    \label{fig:sqrt_s_sigma}
\end{figure*}

At larger $\sqrt{s}$ we fit the $x_F dn/dx_F$ data provided by LHCf \cite{Adriani_2016} in the forward-rapidity region integrated for $p_T<0.4$ at $\sqrt{s}=2.76$ and 7 TeV, where $x_F =2p_z/\sqrt{s}$ is the Feynman-x variable. In particular we consider only the data provided for $x_F<0.7$, since the $x_F$ shape of our $\sigmaInv$ model determined in \us\ is tuned on \cite{2005_NA49} data, which cover $x_F<0.7$ in the low $p_T$ region. 

We have verified a posteriori that the kinematic space of $x_F>0.7$ contributes less than 2\% of the final emissivity described by \ref{eq:source_term}. The highest $\sqrt{s}$ of LHCf is 7 TeV, namely $T_p=2.61 \times 10^7$ GeV in the LAB frame for a fixed target collision. Beyond this incident proton energy our parametrization must be considered as an extrapolation. In the same $\sqrt{s}$ range, data from the ALICE experiment \cite{ALICE:2017nce} are available for the $dn/dy$ at mid-rapidity, and $x_F \sim$0.

Since LHCf provides a larger coverage of the kinematic space, we tune our analysis on this data-set, checking a posteriori that the total multiplicity measured by ALICE is compatible with our result. The inclusion of the ALICE data in the fit would not produce significant differences, being the fit dominated by the LHCf measurements. We perform a $\chi^2$ fit and use the \textsc{Multinest} \cite{Multinest_2009} package to scan over the parameter space. 

Typically, each cross section measurement contains a statistical, a systematic, and a scale uncertainty. For datasets with only a single data point, we can simply add all the individual uncertainties in quadrature. In practice, those are the Demer86 measurements.
We note that the Demer86 data points are a collection from different experiments and therefore have independent uncertainties. 
On the other hand, at higher energies, we use the measurements of $x_F dn/dx_F$ provided by LHCf \cite{Adriani_2016}. For these data points the scaling uncertainty is fully correlated and we cannot simply add them in quadrature in the definition of the total $\chi^2$. Instead, we introduce nuisance parameters allowing for an overall renormalization of each LHCf dataset. We refer to \us\ for a complete explanation of the method, used previously also in ~\cite{Korsmeier_2018}.

Finally, there is one subtlety about the data sets. While the LHCf experiment can distinguish if photons are produced by the $\pi^0$ or an intermediate $\eta$, the collection of measurements in Dermer86 ascribe all photons to the $\pi^0$ decay, namely, they are not corrected for the $\eta$ contribution. Therefore, we correct those data points by subtracting the contributions of $\eta$ using our estimation from Sec. \ref{sec:other_channels}. The
contribution to the total multiplicity varies from $<$0.001\% at $\sqrt{s}=2.2$~GeV to 3\% at $\sqrt{s}=53$~GeV. To be conservative, we increase the uncertainty by adding this correction to the total error in quadrature at each data point.

Overall, our parametrization provides a good fit to the data sets at different $\sqrt{s}$. The $\chi^2$ per number of degrees of freedom (d.o.f.) of the best fit converges to 26/24. The parameters from $c_{20}$ to $c_{27}$ are all well constrained by the fit and their values are reported in Tab.~\ref{tab::Fit_results_pp_pion}. 
The results are displayed in Fig.~ \ref{fig:sqrt_s_sigma}. In the left panel, we report the fit to the low-energy data on the total cross section, while the right panel reports the fit to LHCf data. 
Within our parametrization, the $\pi^0$ total cross section is determined with a precision between 5\% and 10\% below $\sqrt{s}$ of 60~GeV (left panel). At LHCf energies the uncertainty varies between 5\% and 10\% below 0.7 with $x_F$, and increase to more than 10\% for higher $x_F$ values (right panel).
There is a reasonable agreement between our predictions and the data also in the $x_F$ range not considered in the fit. Moreover our model is compatible within $2\sigma$ with the $dn/dy$ measured by ALICE, since it predicts a value of  0.81 at $\sqrt{s}=2.76$ TeV to be compared to $1.803 \pm 0.738$, confirming the goodness of our model.

\begin{table}[b!]
\caption{Results from the best fit and the 1$\sigma$ uncertainty for the parameters of Eq.~\eqref{eq:As}. }
\label{tab::Fit_results_pp_pion}
\begin{tabular}{ l c rll }
 \hline \hline
 Parameter & $\quad$ &  \multicolumn{3}{c}{Best-fit value}  \\
\hline 
$c_{20}$                 &   &   $0.57  $ \!\!\!& $\pm$ &\!\!\!\! $ 0.02$           \\
$c_{21}$                 &   &   $2.25  $ \!\!\!& $\pm$ &\!\!\!\! $ 0.02$  \\
$c_{22}$                 &   &   $-43.33$ \!\!\!& $\pm$ &\!\!\!\! $ 4.54$  \\
$c_{23}$                 &   &   $-6.41 $ \!\!\!& $\pm$ &\!\!\!\! $ 1.14$  \\
$c_{24}$                 &   &   $2.82  $ \!\!\!& $\pm$ &\!\!\!\! $ 0.10$  \\
$c_{25}$                 &   &   $0.06 $ \!\!\!& $\pm$ &\!\!\!\! $ 0.02$  \\
$c_{26}$                 &   &   $27.2  $ \!\!\!& $\pm$ &\!\!\!\! $ 12.6$  \\
$c_{27}$                 &   &   $-0.42  $ \!\!\!& $\pm$ &\!\!\!\! $ 0.09$  \\
 \hline \hline
\end{tabular}
\end{table}

\subsection{Results on the $\gamma$-ray production cross section}\label{sec:gamma_from_pi0}
\label{sec:dsigma_dtgamma}

\begin{figure*}[t]
    \includegraphics[width=0.49\textwidth]{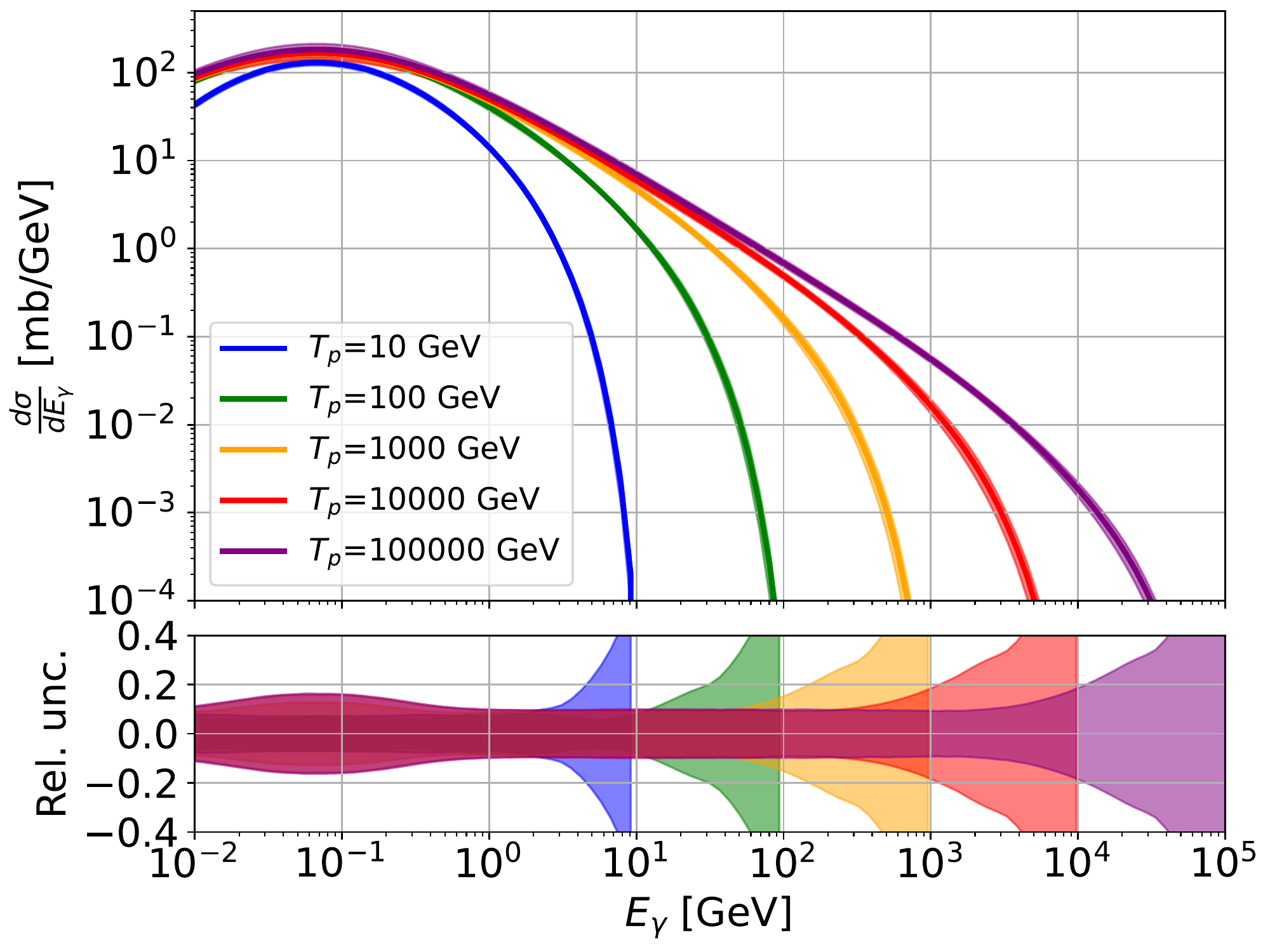}
    \includegraphics[width=0.49\textwidth]{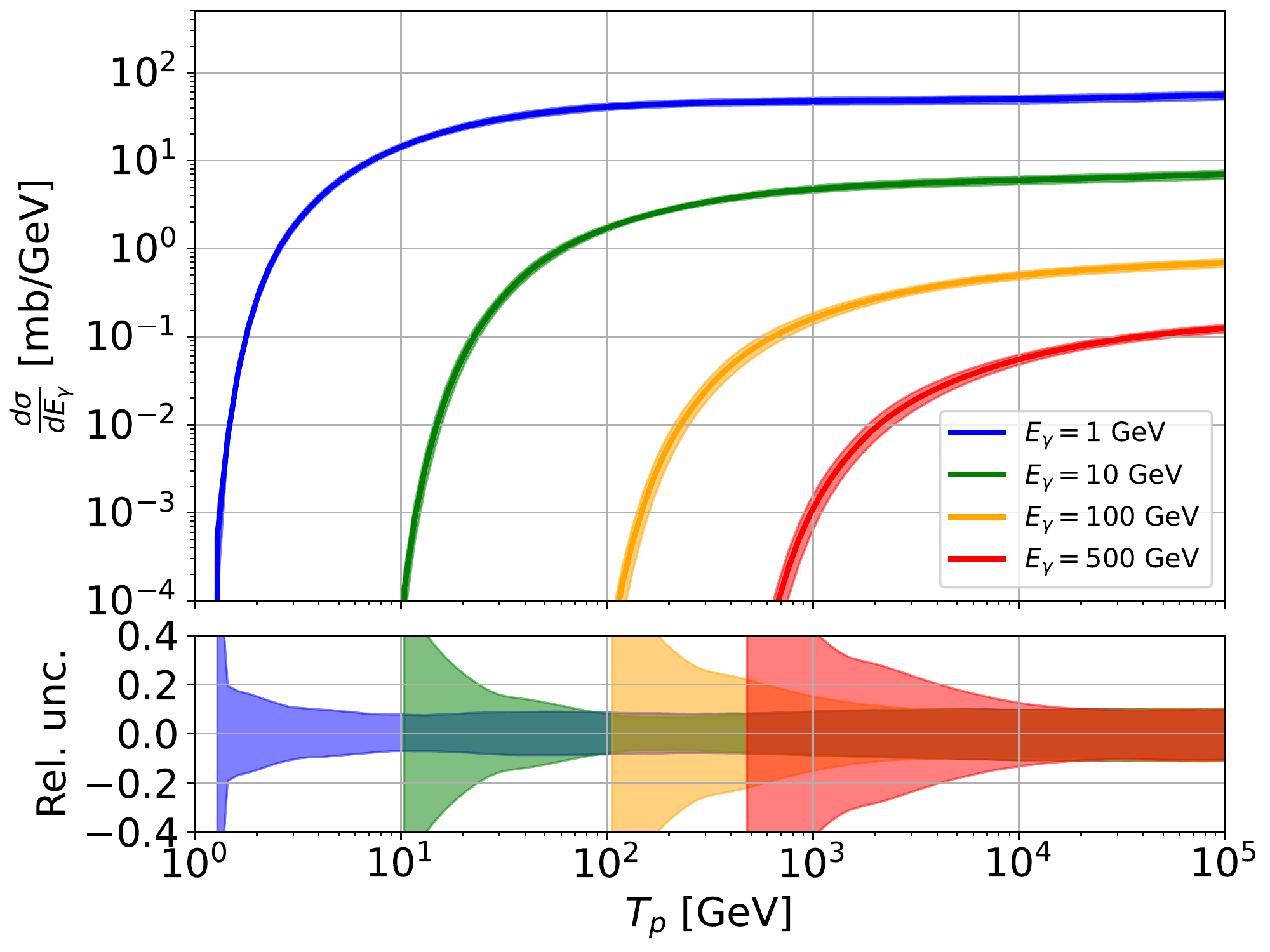}
    \caption{Differential cross section for the production of $\gamma$-rays from $\pi^0$ in $p+p$ collisions, computed for different incident kinetic proton energies as a function of $\gamma$ energy (left) and different $\gamma$ energies as a function of $p$ kinetic energy (right).
    } 
    \label{Fig:pi-zero-final}
\end{figure*}

The differential cross section for the production of $\gamma$-rays from $p+p \rightarrow \pi^0+X$ scattering $d\sigma/d E_{\gamma}$
is obtained from Eq.~\eqref{eq:convolution}, once $\sigmaInv$ is fully determined. 
There are mainly three contributions to the uncertainty band: 
\begin{itemize}
    \item In this work, we have fitted the overall normalization of the $\pi^0$ multiplicity to the Dermer86 and LHCf data using Eq.~\eqref{eq:main_equation}. From the \textsc{MultiNest} scan we obtained the best-fit value and the covariance matrix with correlated uncertainties of the parameters $c_{20}$ to $c_{27}$. We numerically propagate this uncertainty by sampling the cross section parametrization for 500 realizations using the covariance matrix and assuming Gaussian statistics. 
    \item We take the kinetic shape ($G_{\pi^+}$ and $G_{\pi^-}$) from  the previous work \us. These two functions both come with statistical uncertainties. In \us, we derived the covariance matrices for the parameters $c_{1,\pi^+}$ to $c_{19,\pi^+}$, and equivalently for $\pi^-$. This is the statistical uncertainty on the kinematic shape of the cross section. We propagated the uncertainty individually for $\pi^+$ and $\pi^-$, i.e. we assume that the shapes are uncorrelated. Also, the uncertainty of $A(s)$, Eq.~\eqref{eq:main_equation}, is assumed to be uncorrelated from the kinematic shapes. 
    \item Finally, we consider a systematic uncertainty for the kinematic shape. For this, we evaluated the difference of the cross section by assuming either a pure $\pi^+$ or a pure $\pi^-$ kinetic shape. In more detail, it means that in Eq.~\eqref{eq:auxiliary} we replace $G_{\pi^+} + G_{\pi^+}$ by $2G_{\pi^+}$ or $2G_{\pi^-}$, respectively. Then, we derived the energy differential cross section, Eq.~\eqref{eq:convolution}, from these two cases. 
    We compared the two results and use the maximal deviation as a function of energy as an additional contribution to the total uncertainty, which is obtained by adding all contributions in quadrature. 
\end{itemize}

In Fig.~ \ref{Fig:pi-zero-final}, the differential cross section is reported for the production of $\gamma$ from $\pi^0$ in $p+p$ collisions. It is provided for different incident kinetic proton energies $T_p$ as a function of $E_\gamma$ (left) and different $E_\gamma$ energies as a function of $T_p$ (right). Uncertainties are between 6\% and 20\% for most of the energy range, except for $E_{\gamma}$ close to $T_p$, where both statistical and systematic errors increase. For most combinations of $E_\gamma$ and $T_p$ the statistical uncertainty dominates, while the systematic uncertainty due to the kinematic shape is at most at the same level as the statistical error. Only for a region of $E_{\gamma}$ close to $T_p$, which is suppressed in the total emissivity, the systematic uncertainty dominates.
We obtain the most precise prediction for $T_p$ of about 100 GeV, which corresponds to the NA49 and NA61 data for $\pi^\pm$ production.

\section{Contribution from other production channels and from nuclei}
\label{sec:other_channels}

In this section we present our model for the photon production from further intermediate mesons and hyperons, and for scatterings involving nuclei heavier than hydrogen. The decay channels relevant for photon production are:
\begin{itemize}
    \item $K^+\rightarrow\pi^+ \pi^0$ (20.7\%) and $K^+\rightarrow\pi^0 e^+ \nu_e$ (5.1\%), 
    \item $K^-\rightarrow\pi^- \pi^0$ (20.7\%) and $K^-\rightarrow\pi^0 e^- \bar{\nu_e}$ (5.1\%), 
    \item  $K^0_S \rightarrow \pi^0 \pi^0$ ($B_r  =30.7\%$), 
    \item $\Lambda \rightarrow n\pi^0$ ($B_r= 35.8\%$), 
\end{itemize}
where the relevant branching ratio is reported in parenthesis.
We include their contribution by using the production cross sections derived in \us. We calculate the spectra of photons assuming that $\pi^0$ are produced from a two or three body decay. In particular, for three body decays we consider, as in \us, that each of the three particles takes 1/3 of the parent's energy. The $K^0_L$ meson is expected to give a contribution similar to the $K^0_S$ meson in the following decay channels:
\begin{itemize}
    \item $K^0_L \rightarrow \pi^{0}\pi^{0}\pi^{0}$ ($B_r=19.5\%$), 
    \item $K^0_L \rightarrow \pi^{+} \pi^{-} \pi^{0}$ ($B_r=12.5\%$).
\end{itemize}
Due to the lack of experimental data we employ the Pythia event generator \cite{Sjostrand:2014zea} to compare the $p_T$ and $x_F$ dependence of the final photon spectra from $K^0_S$ and $K^0_L$. We find that the $p_T$ and $x_F$ shapes for the production of photons is very similar for the $K^0_L$ and $K^0_S$. The difference is approximately a normalization factor. In particular, the $K^0_L$ meson produces about a factor of 1.16 times more $\pi^0$ than $K^0_S$ which directly translates into an enhancement also for the photon cross section. This is mainly due to the branching ratio of $K^0_L$ into $\pi^0$ which is larger than for $K^0_L$. The ratio between the multiplicity of $\pi^0$ from these two mesons can be calculated as $n_{K^0_L}/n_{K^0_S} = (3\cdot0.195+0.125)/(2\cdot 0.307) \approx 1.16$. In the following we assume that the production cross section of $\gamma$ rays from $K^0_L$ is obtained from $K^0_S$ by a rescaling by a factor $1.16$. 

The hyperons $\bar{\Lambda}$, $\Sigma$ and the $\Xi$ give a subdominant contribution to the total photon yield. For all these particles their pion contributions are usually removed in the data by feed-down corrections. We thus have to add it into our calculations. The multiplicities of $\Omega$ baryons in $p+p$ collisions are a factor of about 3-4 orders of magnitude smaller than the one of $\Lambda$ particles, so we neglect them. We do not consider neither the $\Sigma^-$ particle nor its antiparticle since they both only decay into charged pions.

Since no data are available at the energies of interest, we follow \us\ and  estimate the contribution of the $\bar{\Lambda}$, $\Sigma$ and $\Xi$ baryons using the Pythia code \cite{Sjostrand:2014zea}. In particular, we run the Monte Carlo event generator for $p+p$ collisions in the range $E_p=[20,10^7]$ GeV. We compute the multiplicities $n_{i}$ of each particle $i$, where $i$ runs over $\Sigma^+$, $\Sigma^0$, $\Xi^0$, $\Xi^-$ (and their antiparticles) and $\bar{\Lambda}$. Then, we calculate the ratio $n_i/n_\Lambda$, both derived with Pythia for consistency. Finally, we use the ratio $n_{i}/n_{\Lambda}$ to add these subdominant channels (S.C.) to the total yield of $\gamma$ by rescaling the $\Lambda$ cross sections into a $\gamma$-ray one. Proceeding in this way, we rely on the data-driven invariant cross section of $\Lambda$, which has a comparable mass to all these particles; so we expect the dependence of their cross section with the kinematic parameters to be similar.
Specifically, we use the following prescription:
\begin{equation}
  \frac{d\sigma}{dE_\gamma}(T_p,E_{\gamma})_{\rm S.C.} = \frac{d\sigma}{dE_{\gamma}}(T_p,E_{\gamma})_{\Lambda} \times \sum_i \mathcal{F}^i(T_p),
 \label{eq:sub}
\end{equation}
where $\mathcal{F}^i(T_p)$ represents the correction factor for each particle. For example, for $\Sigma$ particles it can be written as:
\begin{equation}
  \mathcal{F}^{\Sigma}(T_p) =  \frac{n_{\Sigma}(T_p)\cdot {B_r}^{\pi^0}_{\Sigma}} {n_{\Lambda}(T_p)\cdot {B_r}^{\pi^0}_{\Lambda}}
 \label{eq:subfactor}
\end{equation}
where ${B_r}^{\pi^0}_\Sigma$ is the branching ratio for the decay of the $\Sigma$ hyperon into neutral pions.

Below we report the rescaling factor we apply for each particle:
\begin{itemize}

    \item The $\bar{\Lambda}$ decays into $\bar{p}\pi^+$ with $B_r= 63.9\%$ and into $\bar{n}\pi^0$ with $B_r=35.8\%$. Since the branching ratio into $\pi^0$ is the same as for $\Lambda$, the rescaling factor is fixed to $\mathcal{F}^{\Lambda}(T_p)  = n_{\bar{\Lambda}}/n_{\Lambda}$.

    \item The $\Sigma^{+}$ decays with $B_r=51.6\%$ into $p\pi^0$ and $48.4\%$ into $n\pi^+$. Therefore ${B_r}^{\pi^0}_{\Sigma^{+}}=0.52$ and the correction factor $\mathcal{F}$ is given by Eq.~\eqref{eq:subfactor}. Its antiparticle $\bar{\Sigma}^-$ contributes to the photon yield as well.

    \item The $\Sigma^{0}$ decays with $B_r=100\%$ into $\gamma \Lambda$. The correction factor $\mathcal{F}$ is given by $\mathcal{F}^{\Sigma^{0}}= n_{\Sigma^{0}}/n_{\Lambda}$. Its antiparticle, $\bar{\Sigma}^0$, contributes to the photon yield with $\mathcal{F}^{\bar{\Sigma}^{0}}= n_{\bar{\Sigma}^{0}}/n_{\Lambda}$.
    
    \item The $\Xi^0$ decays at almost 100$\%$ into $\pi^0 \Lambda$, so 
    $\mathcal{F}_{\Xi^0} = ((1+Br^{\pi^0}_{\Lambda})\cdot n_{\Xi^0})/(Br^{\pi^0}_{\Lambda} \cdot n_{\Lambda})$. For its antiparticle  $\bar{\Xi}^0$  we take $\mathcal{F}_{\bar{\Xi}^0} = ((1+Br^{\pi^0}_{\bar{\Lambda}})\cdot n_{\bar{\Xi}^0})/(Br^{\pi^0}_{\Lambda} \cdot n_{\Lambda})$.
    
    \item The $\Xi^-$ decays at almost 100$\%$ into $\pi^-\Lambda$. We use for the correction factor in Eq.~\eqref{eq:subfactor} $\mathcal{F}_{\Xi^-} = (Br^{\pi^0}_{\Lambda}\cdot n_{\Xi^-})/(Br^{\pi^0}_{\Lambda} \cdot n_{\Lambda}) = n_{\Xi^-}/n_{\Lambda}$. The antiparticle of $\Xi^-$ is $\bar{\Xi}^+$ which decays into $\pi^+ \bar{\Lambda}$ and has a rescaling factor $\mathcal{F}_{\bar{\Xi}^+} = (Br^{\pi^0}_{\bar{\Lambda}}\cdot n_{\bar{\Xi}^+})/(Br^{\pi^0}_{\Lambda} \cdot n_{\Lambda}) = n_{\bar{\Xi}^+}/n_{\Lambda}$.
\end{itemize}

\begin{figure}[t]
    \includegraphics[width=0.49\textwidth]{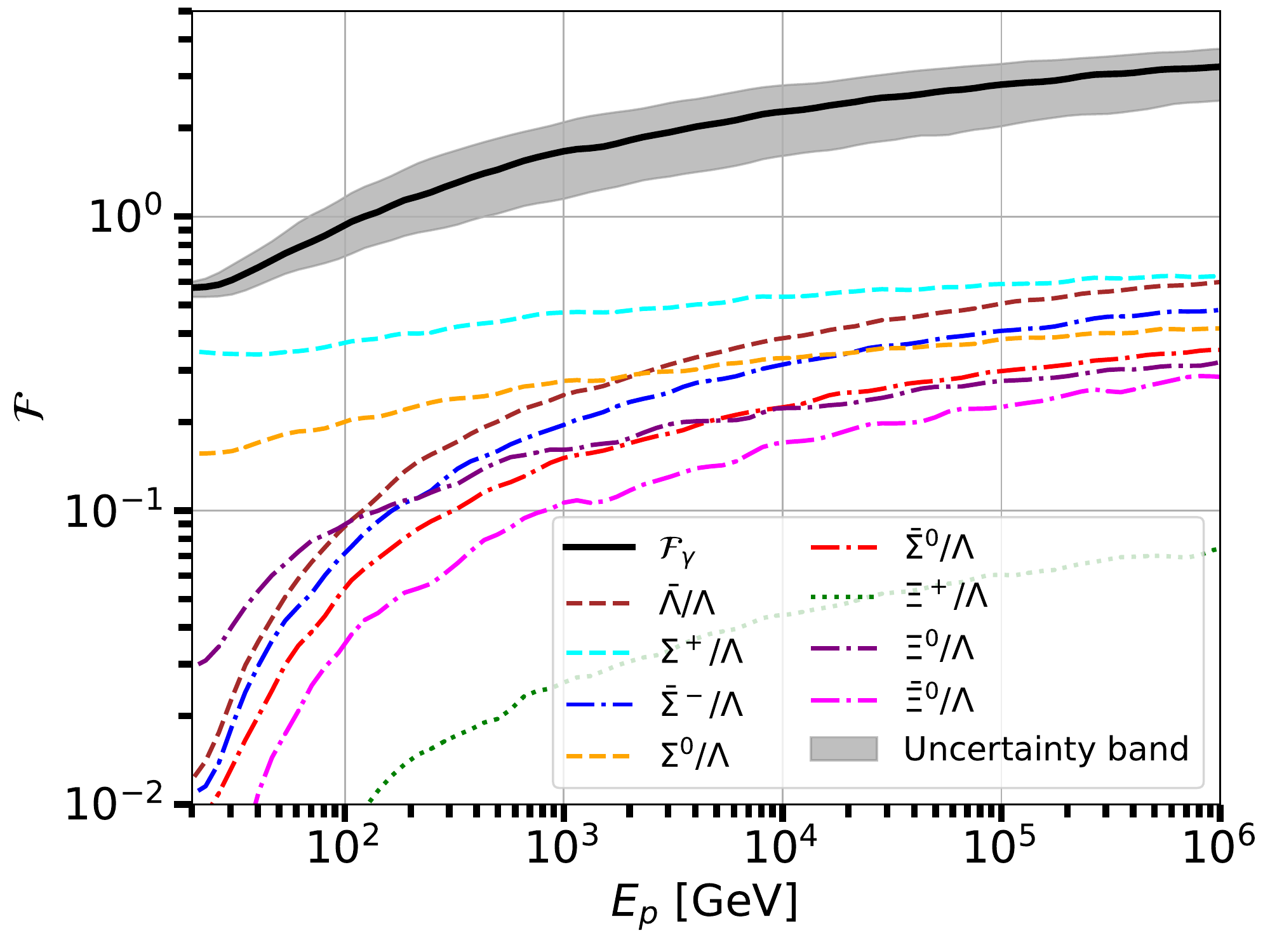}
    \caption{Correction factor $\mathcal{F}$ for the contribution of $\bar{\Lambda}$, $\Sigma$ and $\Xi$ baryons from $p+p$ collisions at different proton energies $E_p$. We show the results for each individual contribution and their sum. We also display the uncertainty band determined by running Pythia with different setup parameters and tunings.} 
    \label{Fig:subdominantchannels}
\end{figure}

In Fig.~\ref{Fig:subdominantchannels}, we report the correction factor $\mathcal{F}$ for the subdominant channels that contribute to photon yield.
At low energy, $\mathcal{F}$ is between $40\%$ and $100\%$ while at high energy it reaches a factor of 3.
We also show the variation to $\mathcal{F}$ obtained from different Pythia setups (uncertainty band, see \us\ for more details).

\begin{figure}
    \includegraphics[width=0.49\textwidth]{"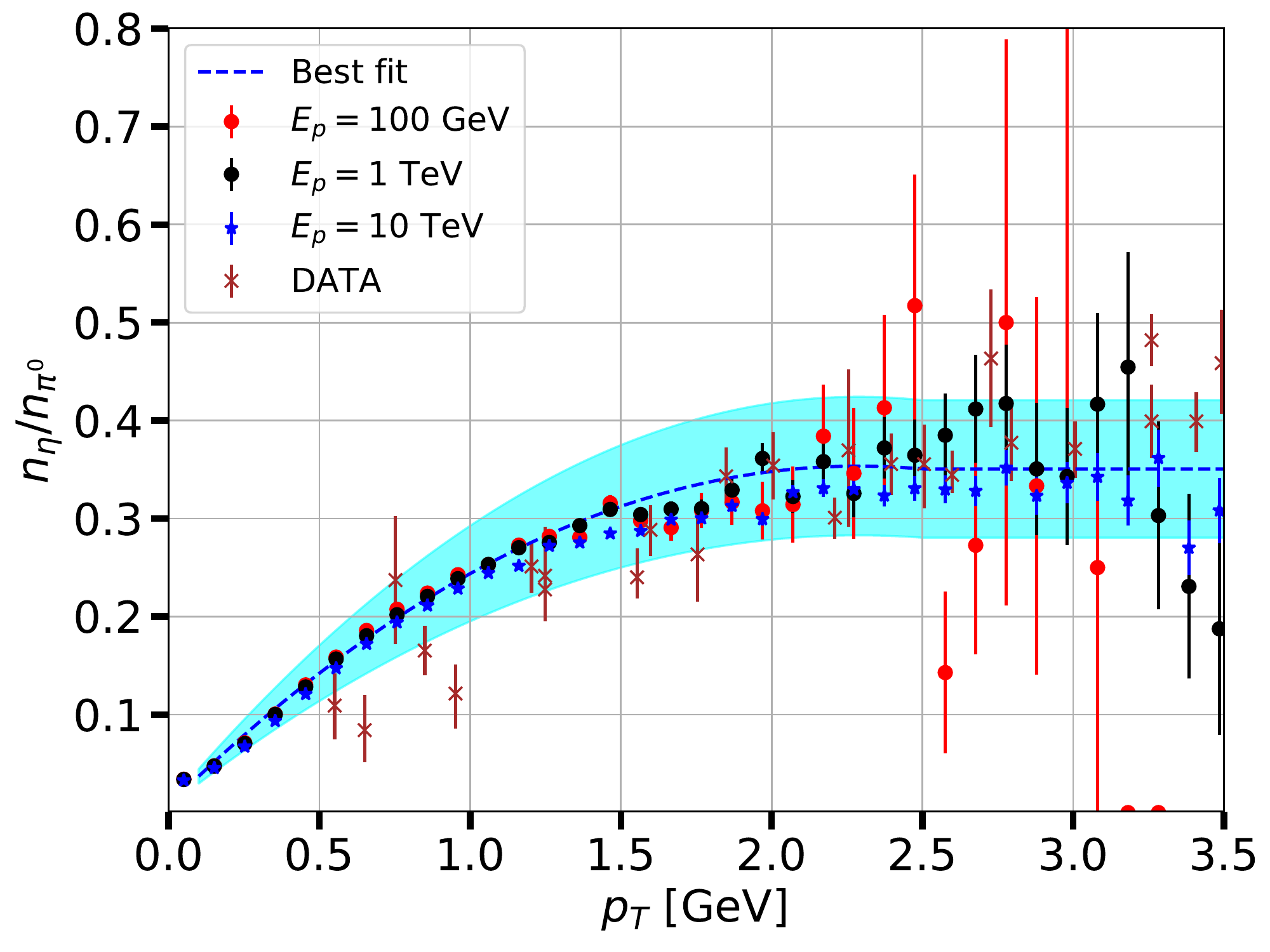"}
  \caption{Ratio between the multiplicity of $\eta$ meson and $\pi^0$ as a function of the transverse momentum obtained with Pythia using different incoming proton energies (red, blue, and black data points). We also show the data (brown data points) and the best fit and uncertainty band obtained with our model.}
  \label{fig:etaprod}
\end{figure}

Another relevant channel for the production of photons is the $\eta$ meson, which decays into: 
\begin{itemize}
    \item $\eta \rightarrow \gamma \gamma$ ($B_r= 39.41\%$),
    \item $\eta \rightarrow \pi^0 \pi^0 \pi^0$ ($B_r= 32.68\%$),
    \item $\eta \rightarrow \pi^+ \pi^- \pi^0$ ($B_r= 22.92\%$),
    \item $\eta \rightarrow \pi^+ \pi^- \gamma$ ($B_r= 4.22\%$).
\end{itemize}
Cross section data for the production of $\eta$ mesons have been recently measured by the ALICE experiment at $\sqrt{s} = 2.76$, 7 and 8 TeV \cite{ALICE:2017ryd,ALICE:2017nce}, and by PHENIX at $\sqrt{s} = 200$ GeV \cite{PHENIX:2010hvs}. Older measurements are reported in Ref.~\cite{WA80:1995whm} and references therein. These data are typically collected at mid-rapidity and the double differential cross section data is not available. The $\pi^0$ produced in the second and third decay channels are not distinguished experimentally from the prompt ones because the decay time of $\eta$ is much smaller than the one of $\pi^0$. Therefore, the $\pi^0$ production from $\eta$ decay is already included in the total one as described in previous Section. We include the photons from the direct $\eta$ decay ($\eta \rightarrow \gamma \gamma$) by using the measured ratio between its multiplicity with respect $\pi^0$ one, as a function of $p_T$.
This is measured for $p_T$ from 0.5 to 5 GeV and shows an increasing trend, as visible in Fig.~ \ref{fig:etaprod}. 
Since $\eta$ and $\pi^0$ mesons have different branching ratios for the direct decay into two photons, experimentally the multiplicity of the process $\eta \rightarrow \gamma \gamma$ has been rescaled for $1/B_r$, where $B_r$ is the branching ratio of this process ($1/0.3941$).
At low $p_T$ the ratio between the $\eta$ and $\pi^0$ multiplicites $n_{\eta}/n_{\pi^0}$ is of the order of 0.05-0.15 while at high $p_T$ reaches a plateau at the level of 0.4. Some of the measurements for $n_{\eta}/n_{\pi^0}$ are at mid rapidity, e.g.~for the ALICE experiment \cite{ALICE:2017ryd,ALICE:2017nce}, while others are integrated over a different range of kinematic variables. 
Since most of the contribution to the $\gamma$-ray source term is at low $p_T$ we expect the contribution of $\eta$  to be at the level of $5-15\% \times Br(\eta \rightarrow \gamma \gamma) = 2-6\%$. We report in Fig.~\ref{fig:etaprod} the results obtained with the simulations of Pythia together with the model that reproduces well both the simulations and the data. We use for this scope a function with different powers of $p_T$. The contribution from the forth channel only contributes less then 0.5\% and thus it is neglected.

As for the inclusion of scatterings including nuclei, in either the CRs or in the ISM, we closely follow the prescriptions derived in ODDK22 for $\pi^{\pm}$, given the lack of any dedicated data.  Specifically, if a $\pi^0$ is produced in collisions between projectile and target nuclei with  $A_1$ and $A_2$ mass numbers, the $G$ functions in Eq.~ \ref{eq:main_equation} are corrected as in Eqs.~(25)--(27) by \us. The parameters in Eq.~(26) are taken from Tab.~V from \us, where column $\pi^+$ ($\pi^-$) corrects the function $G_{\pi^+}$ ($G_{\pi^-}$). The $K^\pm$ channel is modified analogously by using the columns 3 and 4 in  Tab.~V from \us. For all the other channels, we assume a correction function which is the average from the $K^+$ and $K^-$ ones.

\section{Results on the $\gamma$ ray production cross section and emissivity}
\label{sec:results}

\begin{figure*}[t]
    \includegraphics[width=0.49\textwidth]{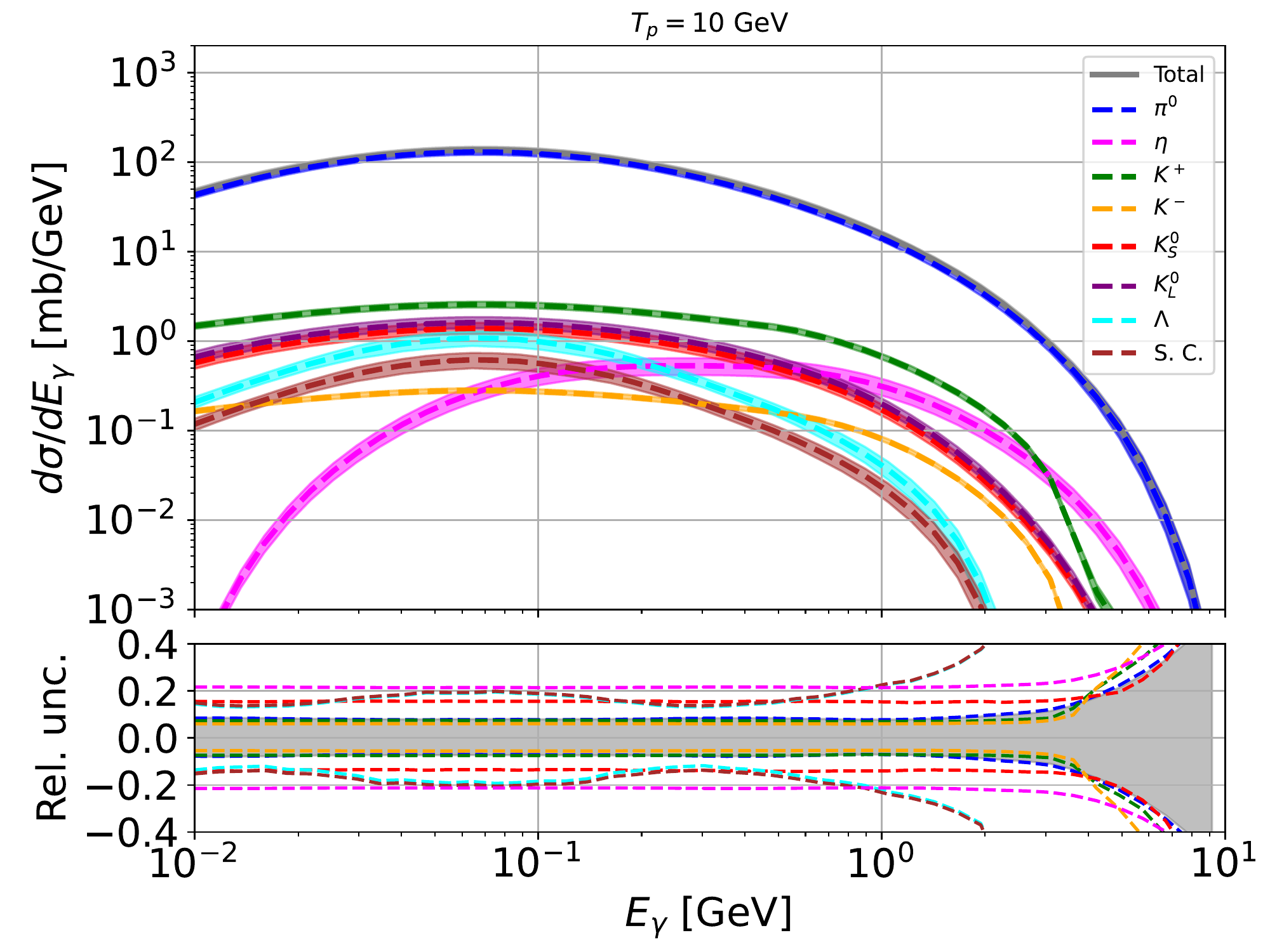}
    \includegraphics[width=0.49\textwidth]{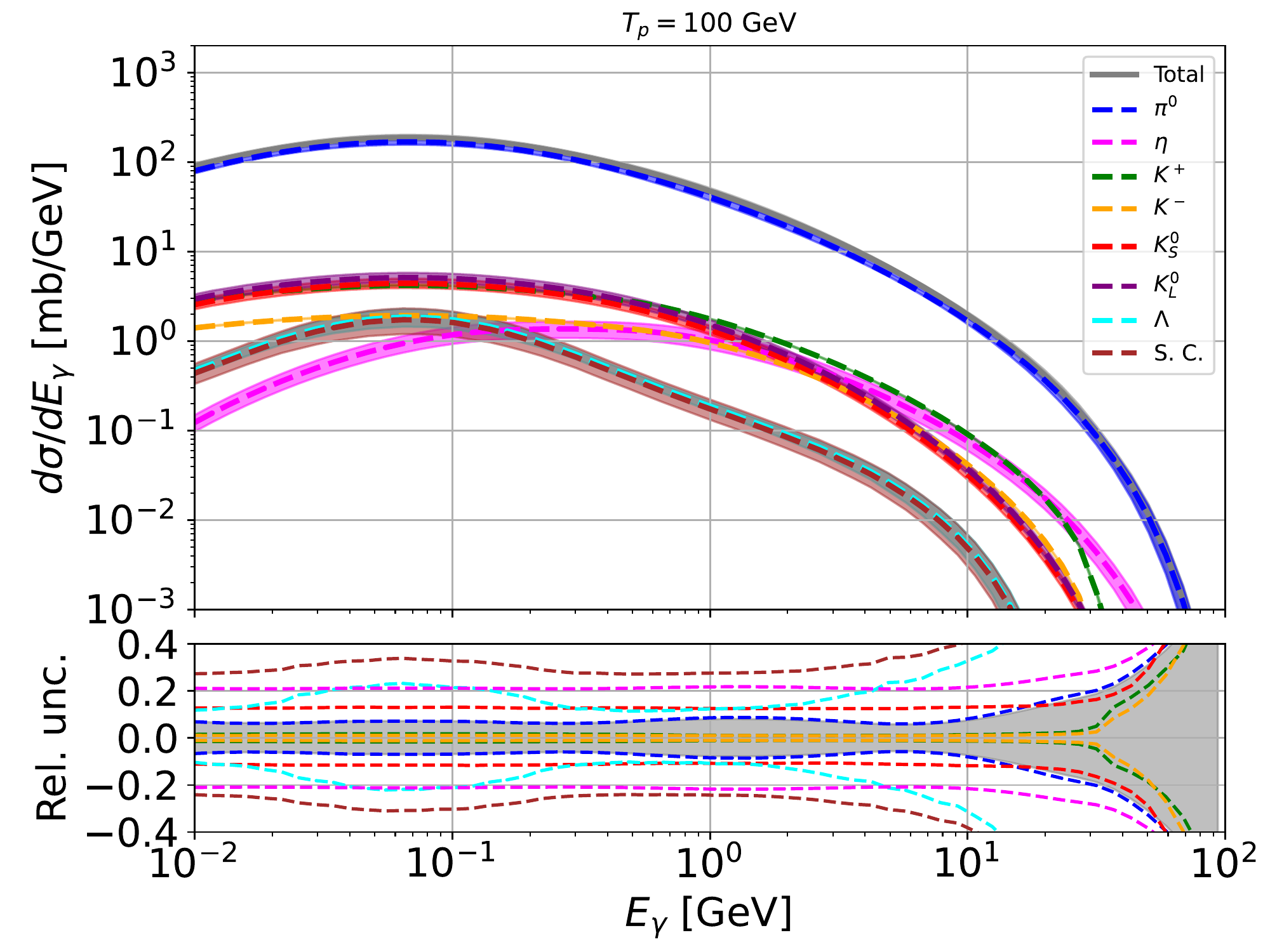}
    \includegraphics[width=0.49\textwidth]{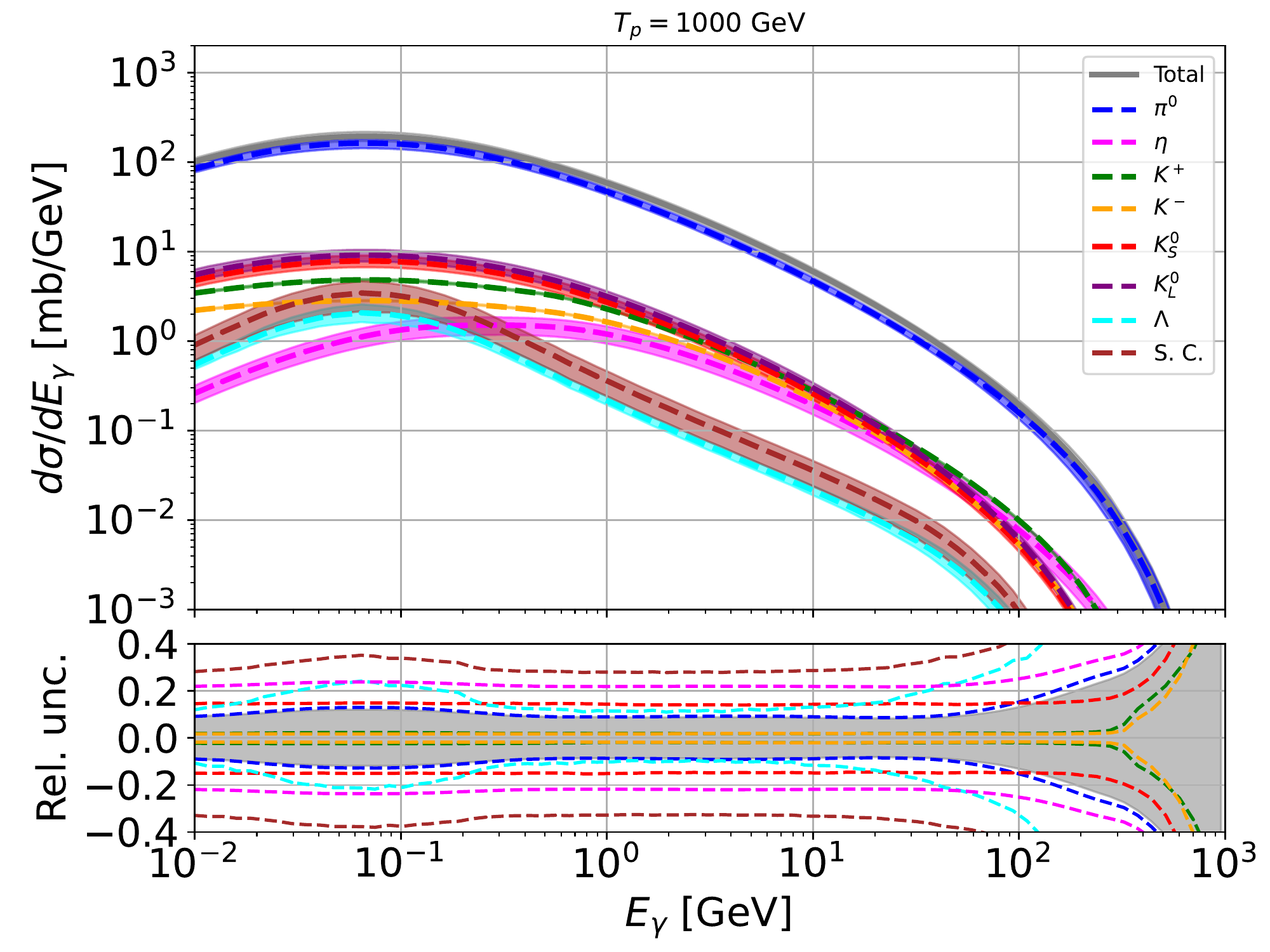}
    \includegraphics[width=0.49\textwidth]{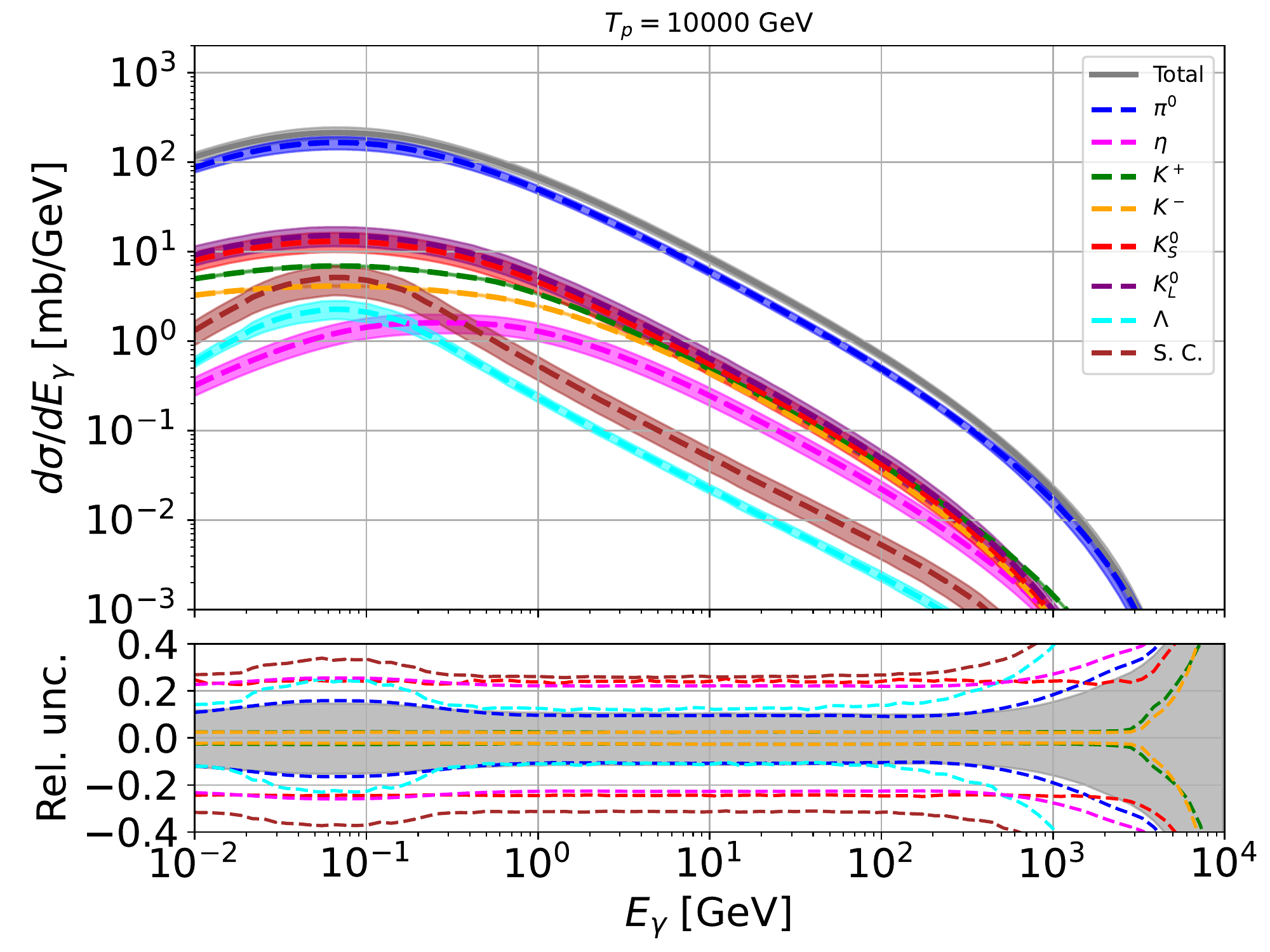}
    \caption{Differential cross section for the inclusive production of $\gamma$ in $p+p$ collisions, derived from fits to the data as described in Secs. \ref{sec:pi0} and \ref{sec:other_channels}. We plot separate production of $\pi^0$,$\eta$, $K^+$, $K^-$, $K_0^S$, $K_0^L$, $\Lambda$  and subdominant channels (S.C.), and their sum. Each plot is computed for incident proton energies $T_p$ of 10, 100, 1000 and 10000 GeV. The curves are displayed along with their 1$\sigma$ error band. At the bottom of each panel the 1$\sigma$ uncertainty band is displayed around the best fit individually for each contribution.} 
    \label{Fig:cross-final}
\end{figure*}

We now can compute the total differential cross section $d\sigma/dE_{\gamma}$ for the inclusive production of $\gamma$ rays in $p+p$ inelastic collisions. The result is obtained by summing all the contributions from $\pi^0$ and the subdominant channels, as discussed in Secs.~\ref{sec:pi0} and \ref{sec:other_channels}. This is the main result of our paper and it is shown in Fig.~\ref{Fig:cross-final} for four representative incident proton energies. The contribution of $\pi^0 \rightarrow 2 \gamma$ is dominant at all proton and photon energies. 

For the decays of $\eta$, $K^+$, $K^-$, $K_0^S$, $K_0^L$, and $\Lambda$ we also show the individual contributions, while all the subdominant channels are combined into a single curve. All these channels contribute at most few percent of the total cross section. However, their shapes, as a function of $T_p$ and $E_\gamma$, slightly differ from the dominant $\pi^0$ channel. The gray curve and shaded band display the total $d\sigma/dE_{\gamma}$ and the $1\sigma$ uncertainty band, respectively. The final uncertainty spans from 6\% to 20\% at different $T_p$ and $E_\gamma$, and is driven by the modeling of the $\pi^0$ cross section. As already specified,  the highest $\sqrt{s}$ of LHCf is 7 TeV corresponds to $T_p=2.61 \times 10^7$ GeV in the LAB frame for a fixed target collision, as the ones occurring in the Galaxy. Beyond this limit, our parametrizations are not validated on data, and their values must be considered as an extrapolation.

\begin{figure*}
    \begin{minipage}[t]{0.45\textwidth}
    \includegraphics[width=1.0\textwidth]{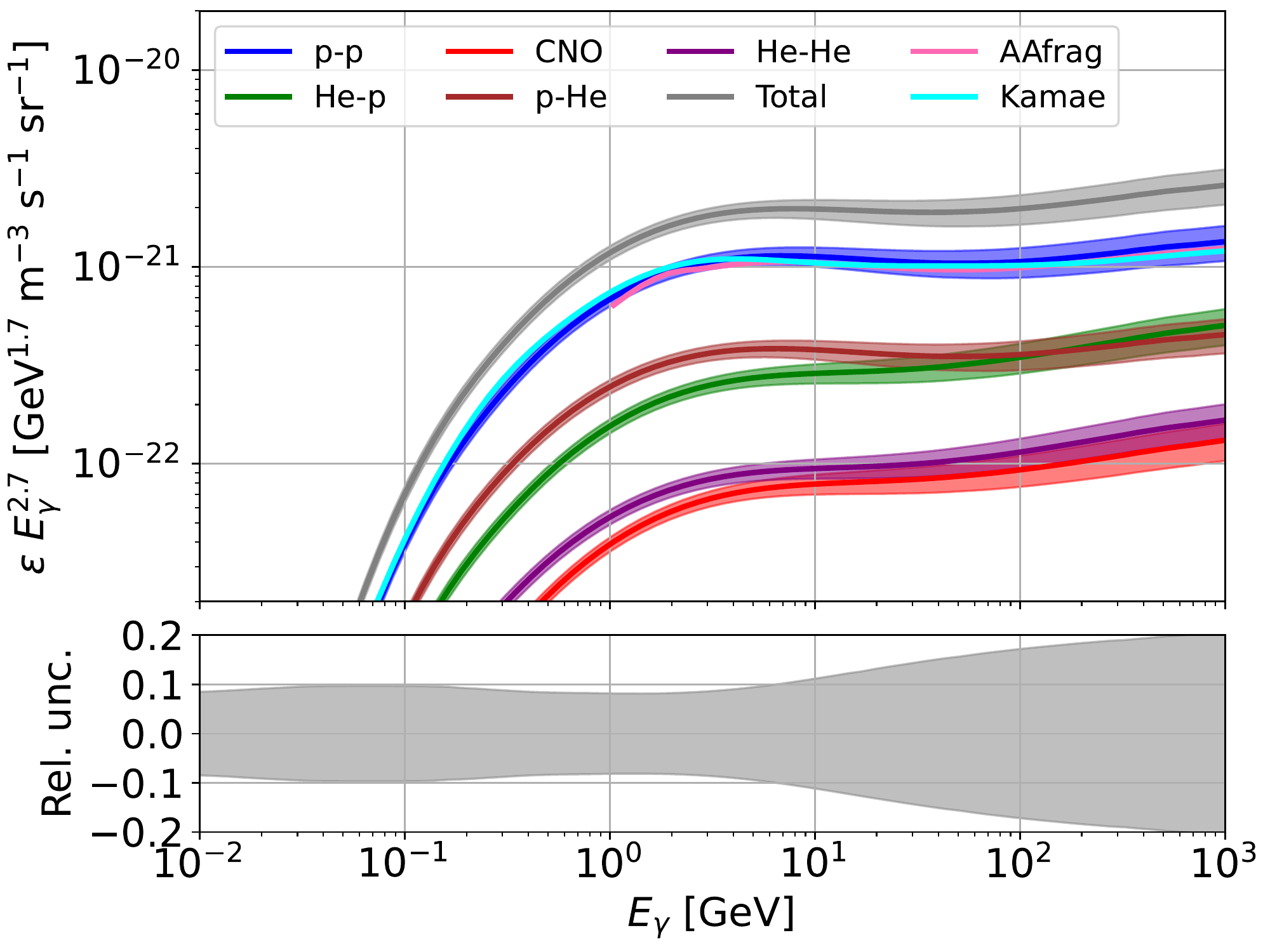}    
    \caption{The $\gamma$-ray emissivity is computed for $p+p$, He$+p$, $p+$He, He+He and CNO$+p$ scatterings. The grey line is the sum of all contributions (see text for details). Each prediction is plotted with the relevant uncertainty due to the production cross section derived in this paper. In the bottom panel, the relative uncertainty to the total $\epsilon(E_\gamma)$ is reported. For comparison, we show the results by \cite{Kamae:2006bf} (Kamae) and \cite{Koldobskiy:2021nld} (AAfrag) in the $p+p$ channel.
   } 
    \label{Fig:emissivity}
    \end{minipage}\hspace{0.04\textwidth}\begin{minipage}[t]{0.45\textwidth}
        \includegraphics[width=1.0\textwidth]{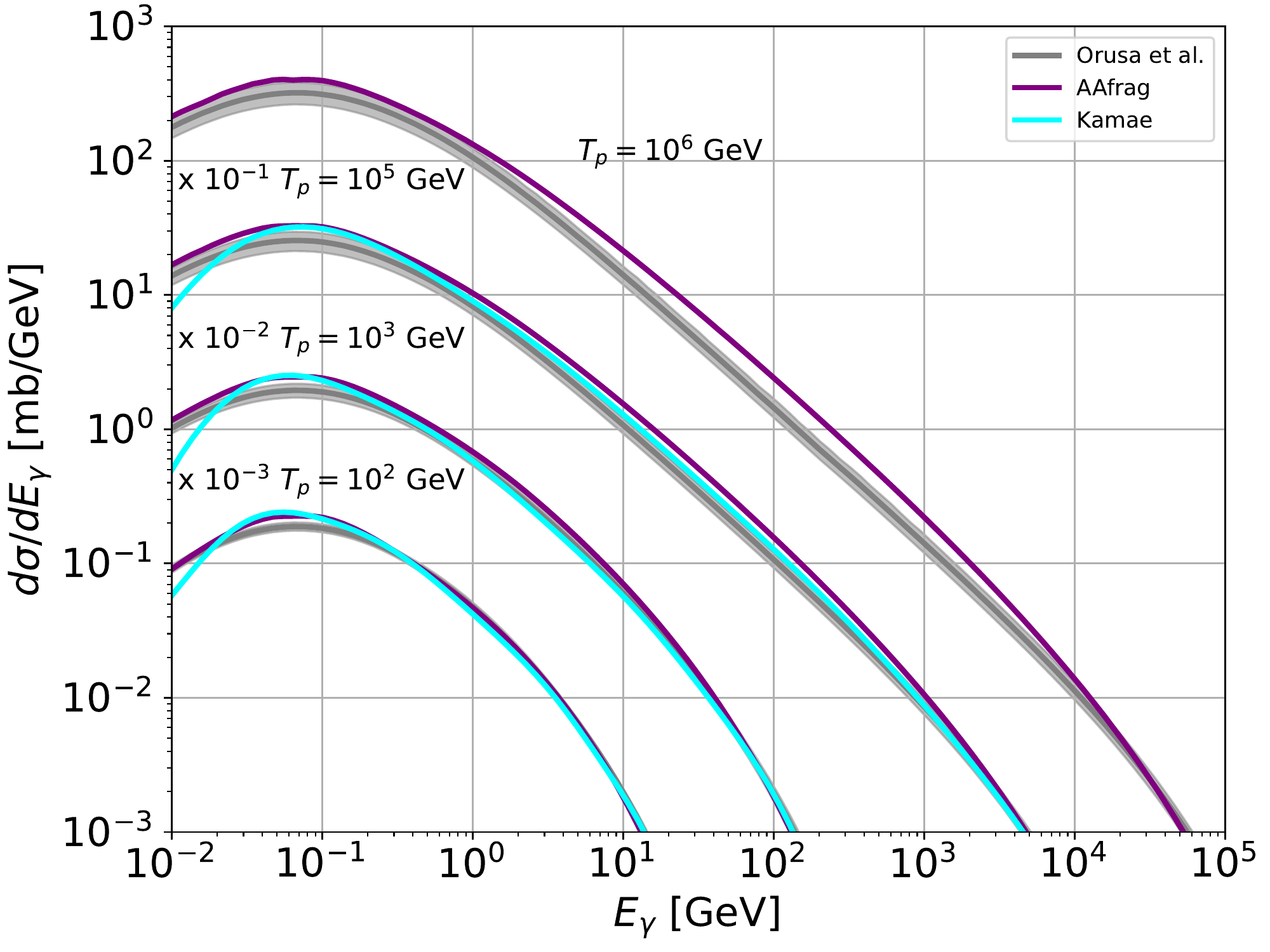}    
    \caption{Comparison among our differential cross section and the one reported in \cite{Kamae:2006bf} (Kamae) and in \cite{Koldobskiy:2021nld} (AAfrag), for incident proton energy $T_p =
    10^6, 10^5, 10^3$ and $ 10^2$ GeV. The three lower curves have been rescaled by the factor indicated in the figure for the sake of visibility.
   } 
    \label{Fig:qE_high_energy_comparison}
    \end{minipage}
\end{figure*}

For illustration, we compute the emissivity in Eq.~\ref{eq:source_term} assuming a constant $n_{\rm ISM}$ and incident CR spectra independent of Galactic position. In Fig.~\ref{Fig:emissivity} we show $\epsilon(E_\gamma)$  as a function of $E_\gamma$ for $p+p$, He$+p$, $p$+He, He+He and CNO$+p$ scatterings, and their sum. We assume $n_{\rm H} = 0.9 \, {\rm cm}^{-3}$ and $n_{\rm He} = 0.1 \, {\rm cm}^{-3}$. 
Each prediction is plotted with the relevant uncertainty from the production cross section derived in this paper. The relative uncertainty to the total $\epsilon(E_\gamma)$ is reported in the bottom panel. As expected, the most relevant contribution comes for $p+p$ reactions. Nevertheless, the contributions from scatterings involving helium globally produce a comparable source spectrum. The uncertainty on $\epsilon(E_\gamma)$ due to hadronic production cross sections is about 10\% for $E_\gamma \leq 10 $ GeV, and increases  to 20\% at TeV energies.  As a comparison, we report the results by \cite{Kamae:2006bf} (Kamae) and \cite{Koldobskiy:2021nld} (AAfrag) for the $p+p$ channel. The latter is plotted for $E_\gamma  > 1 $ GeV since their results start from $T_p > 4$ GeV. 

In order to estimate the impact of our results on the diffuse Galactic emission at {\em Fermi}-LAT energies, we show 
in Fig.~\ref{Fig:qE_high_energy_comparison} a comparison between our cross section and the one derived by Kamae et al.~\cite{Kamae:2006bf} that is used in the {\em Fermi}-LAT official Galactic interstellar emission model \cite{Fermi-LAT:2016zaq}. We also report the results obtained with AAfrag \cite{Koldobskiy:2021nld}. 

As a general comment, our cross section is larger than Kamae et al. at  {\em Fermi}-LAT energies by a rough 5-10\%, depending on the energies. Also, the high energy trend of our cross section is slightly harder than Kamae and AAfrag. As an example, for $T_p=1$ TeV, the Kamae  cross section is lower than ours by $(-1;-5;-12) \%$ at $E_\gamma = (1;50;200)$ GeV. The difference of our result with the AAfrag cross section is $(+16;+5;-15) \%$ at same $E_\gamma$.
The emissivity shown in Fig. \ref{Fig:emissivity}  is comparable or slightly higher with respect to the Kamae and AAfrag ones. Fig. \ref{Fig:qE_high_energy_comparison} shows how our model predicts similar or slightly higher values of the cross-section for those $E_{\gamma}$ produced in the forward direction, that are the relevant ones for the emissivity in the plotted energy range.
In the relevant energies for {\em Fermi}-LAT, the results obtained in this paper are however compatible with Kamae and AAfrag at 1$\sigma$ of the estimated uncertainty bands.

\section{Discussion and conclusions}
\label{sec:conclusions}

The secondary production of $\gamma$ rays from hadronic collisions is a major source of energetic photons in the Galaxy. The diffuse Galactic emission is dominated by the decay of neutral pions, in turns produced by the inelastic scattering of nuclei CRs with the ISM. A precise modeling of the production cross section of $\gamma$ rays of hadronic origin is crucial for the interpretation of data coming from the {\em Fermi}-LAT, for which the diffuse emission is an unavoidable foreground to any source or diffuse data analysis. In the near future, the full exploitation of the data from CTA is subject to a deep understanding of the diffuse emission.

In this paper, we propose a new evaluation for the production cross section of $\gamma$ rays from $p+p$ collisions, employing the scarce existing data on the total cross sections, and relying on previous analysis of the cross section for $e^\pm$. We consider all the production channels contributing at least to 0.5\% level. The cross section for scattering of nuclei heavier than protons is also derived. Our results are supplied by a realistic and conservative estimation of the uncertainties affecting the differential cross section $d\sigma/dE_{\gamma}$, intended as the sum of all the production channels. This cross section is estimated here with an error of 10\% for $E_\gamma \leq 10$ GeV, increasing to 20\% at 1 TeV. 

We also provide a comparison with the cross sections implemented in the official model for the {\em Fermi}-LAT diffuse emission from hadronic scatterings. It turns out that our cross section  is higher than the one in \cite{Kamae:2006bf} by an average 10\% , depending on impinging protons and $\gamma$-ray energies. This result is relevant for the {\em Fermi}-LAT data analysis in the regions close to the Galactic plane, where hadronic scatterings with ISM nuclei are the main source of diffuse photons. 

In order to improve the accuracy of the present result, new data from colliders are needed. Specifically, data is  required on the Lorentz invariant cross section, and not only on the total cross section, for $\pi^0$ productions. The most important kinetic parameter space is $p_T \lesssim 1$ GeV, a large coverage in $x_R$ and beam energies in the LAB frame covering from a few tens of GeV to at least a few TeV. It would be important to get the same measurements also on a He target. Being interested in the $\gamma$ rays produced in the Galaxy, it would also be practical to have data on the inclusive $\gamma$-ray production cross section, and not only on the individual channels. 

We provide numerical tables for the energy-differential cross sections $d\sigma/dE_{\gamma}$ as a function of the $E_{\gamma}$ and incident proton (nuclei) energies from 0.1 to $10^7$ GeV (GeV/n), and a script to read them. 
The material is available at \url{https://github.com/lucaorusa/gamma_cross_section}.

\section*{Acknowledgments} 

MDM research is supported by Fellini - Fellowship for Innovation at INFN, funded by the European Union’s Horizon 2020 research program under the Marie Skłodowska-Curie Cofund Action, grant agreement no.~754496.
FD and LO acknowledge the support the Research grant {\sc TAsP} (Theoretical Astroparticle Physics) funded by Istituto Nazionale di Fisica Nucleare. LO  has been partially supported by ASI (Italian Space Agency) and CAIF (Cultural Association of Italians at Fermilab).
MK is supported by the Swedish Research Council under contracts 2019-05135 and 2022-04283 and the European Research Council under grant 742104.

\bibliography{paper}

\begin{thebibliography}{55}%
\makeatletter
\providecommand \@ifxundefined [1]{%
 \@ifx{#1\undefined}
}%
\providecommand \@ifnum [1]{%
 \ifnum #1\expandafter \@firstoftwo
 \else \expandafter \@secondoftwo
 \fi
}%
\providecommand \@ifx [1]{%
 \ifx #1\expandafter \@firstoftwo
 \else \expandafter \@secondoftwo
 \fi
}%
\providecommand \natexlab [1]{#1}%
\providecommand \enquote  [1]{``#1''}%
\providecommand \bibnamefont  [1]{#1}%
\providecommand \bibfnamefont [1]{#1}%
\providecommand \citenamefont [1]{#1}%
\providecommand \href@noop [0]{\@secondoftwo}%
\providecommand \href [0]{\begingroup \@sanitize@url \@href}%
\providecommand \@href[1]{\@@startlink{#1}\@@href}%
\providecommand \@@href[1]{\endgroup#1\@@endlink}%
\providecommand \@sanitize@url [0]{\catcode `\\12\catcode `\$12\catcode
  `\&12\catcode `\#12\catcode `\^12\catcode `\_12\catcode `\%12\relax}%
\providecommand \@@startlink[1]{}%
\providecommand \@@endlink[0]{}%
\providecommand \url  [0]{\begingroup\@sanitize@url \@url }%
\providecommand \@url [1]{\endgroup\@href {#1}{\urlprefix }}%
\providecommand \urlprefix  [0]{URL }%
\providecommand \Eprint [0]{\href }%
\providecommand \doibase [0]{http://dx.doi.org/}%
\providecommand \selectlanguage [0]{\@gobble}%
\providecommand \bibinfo  [0]{\@secondoftwo}%
\providecommand \bibfield  [0]{\@secondoftwo}%
\providecommand \translation [1]{[#1]}%
\providecommand \BibitemOpen [0]{}%
\providecommand \bibitemStop [0]{}%
\providecommand \bibitemNoStop [0]{.\EOS\space}%
\providecommand \EOS [0]{\spacefactor3000\relax}%
\providecommand \BibitemShut  [1]{\csname bibitem#1\endcsname}%
\let\auto@bib@innerbib\@empty
\bibitem [{\citenamefont {Atwood}\ \emph {et~al.}(2009)\citenamefont {Atwood}
  \emph {et~al.}}]{Fermi-LAT:2009ihh}%
  \BibitemOpen
  \bibfield  {author} {\bibinfo {author} {\bibfnamefont {W.~B.}\ \bibnamefont
  {Atwood}} \emph {et~al.} (\bibinfo {collaboration} {Fermi-LAT}),\ }\href
  {\doibase 10.1088/0004-637X/697/2/1071} {\bibfield  {journal} {\bibinfo
  {journal} {Astrophys. J.}\ }\textbf {\bibinfo {volume} {697}},\ \bibinfo
  {pages} {1071} (\bibinfo {year} {2009})},\ \Eprint
  {http://arxiv.org/abs/0902.1089} {arXiv:0902.1089 [astro-ph.IM]} \BibitemShut
  {NoStop}%
\bibitem [{\citenamefont {Abdollahi}\ \emph {et~al.}(2020)\citenamefont
  {Abdollahi} \emph {et~al.}}]{Fermi-LAT:2019yla}%
  \BibitemOpen
  \bibfield  {author} {\bibinfo {author} {\bibfnamefont {S.}~\bibnamefont
  {Abdollahi}} \emph {et~al.} (\bibinfo {collaboration} {Fermi-LAT}),\ }\href
  {\doibase 10.3847/1538-4365/ab6bcb} {\bibfield  {journal} {\bibinfo
  {journal} {Astrophys. J. Suppl.}\ }\textbf {\bibinfo {volume} {247}},\
  \bibinfo {pages} {33} (\bibinfo {year} {2020})},\ \Eprint
  {http://arxiv.org/abs/1902.10045} {arXiv:1902.10045 [astro-ph.HE]}
  \BibitemShut {NoStop}%
\bibitem [{\citenamefont {Ballet}\ \emph {et~al.}(2020)\citenamefont {Ballet},
  \citenamefont {Burnett}, \citenamefont {Digel},\ and\ \citenamefont
  {Lott}}]{Ballet:2020hze}%
  \BibitemOpen
  \bibfield  {author} {\bibinfo {author} {\bibfnamefont {J.}~\bibnamefont
  {Ballet}}, \bibinfo {author} {\bibfnamefont {T.~H.}\ \bibnamefont {Burnett}},
  \bibinfo {author} {\bibfnamefont {S.~W.}\ \bibnamefont {Digel}}, \ and\
  \bibinfo {author} {\bibfnamefont {B.}~\bibnamefont {Lott}} (\bibinfo
  {collaboration} {Fermi-LAT}),\ }\href@noop {} {\  (\bibinfo {year} {2020})},\
  \Eprint {http://arxiv.org/abs/2005.11208} {arXiv:2005.11208 [astro-ph.HE]}
  \BibitemShut {NoStop}%
\bibitem [{\citenamefont {Abdollahi}\ \emph {et~al.}(2022)\citenamefont
  {Abdollahi} \emph {et~al.}}]{Fermi-LAT:2022byn}%
  \BibitemOpen
  \bibfield  {author} {\bibinfo {author} {\bibfnamefont {S.}~\bibnamefont
  {Abdollahi}} \emph {et~al.} (\bibinfo {collaboration} {Fermi-LAT}),\ }\href
  {\doibase 10.3847/1538-4365/ac6751} {\bibfield  {journal} {\bibinfo
  {journal} {Astrophys. J. Supp.}\ }\textbf {\bibinfo {volume} {260}},\
  \bibinfo {pages} {53} (\bibinfo {year} {2022})},\ \Eprint
  {http://arxiv.org/abs/2201.11184} {arXiv:2201.11184 [astro-ph.HE]}
  \BibitemShut {NoStop}%
\bibitem [{\citenamefont {Aleksi\'c}\ \emph {et~al.}(2016)\citenamefont
  {Aleksi\'c} \emph {et~al.}}]{MAGIC:2014zas}%
  \BibitemOpen
  \bibfield  {author} {\bibinfo {author} {\bibfnamefont {J.}~\bibnamefont
  {Aleksi\'c}} \emph {et~al.} (\bibinfo {collaboration} {MAGIC}),\ }\href
  {\doibase 10.1016/j.astropartphys.2015.02.005} {\bibfield  {journal}
  {\bibinfo  {journal} {Astropart. Phys.}\ }\textbf {\bibinfo {volume} {72}},\
  \bibinfo {pages} {76} (\bibinfo {year} {2016})},\ \Eprint
  {http://arxiv.org/abs/1409.5594} {arXiv:1409.5594 [astro-ph.IM]} \BibitemShut
  {NoStop}%
\bibitem [{\citenamefont {De~Naurois}(2020)}]{DeNaurois:2020bac}%
  \BibitemOpen
  \bibfield  {author} {\bibinfo {author} {\bibfnamefont {M.}~\bibnamefont
  {De~Naurois}} (\bibinfo {collaboration} {H.E.S.S.}),\ }\href {\doibase
  10.22323/1.358.0656} {\bibfield  {journal} {\bibinfo  {journal} {PoS}\
  }\textbf {\bibinfo {volume} {ICRC2019}},\ \bibinfo {pages} {656} (\bibinfo
  {year} {2020})}\BibitemShut {NoStop}%
\bibitem [{\citenamefont {Acharya}\ \emph {et~al.}(2013)\citenamefont {Acharya}
  \emph {et~al.}}]{CTAConsortium:2013ofs}%
  \BibitemOpen
  \bibfield  {author} {\bibinfo {author} {\bibfnamefont {B.~S.}\ \bibnamefont
  {Acharya}} \emph {et~al.} (\bibinfo {collaboration} {CTA Consortium}),\
  }\href {\doibase 10.1016/j.astropartphys.2013.01.007} {\bibfield  {journal}
  {\bibinfo  {journal} {Astropart. Phys.}\ }\textbf {\bibinfo {volume} {43}},\
  \bibinfo {pages} {3} (\bibinfo {year} {2013})}\BibitemShut {NoStop}%
\bibitem [{\citenamefont {Albert}\ \emph {et~al.}(2020)\citenamefont {Albert}
  \emph {et~al.}}]{HAWC:2020hrt}%
  \BibitemOpen
  \bibfield  {author} {\bibinfo {author} {\bibfnamefont {A.}~\bibnamefont
  {Albert}} \emph {et~al.} (\bibinfo {collaboration} {HAWC}),\ }\href {\doibase
  10.3847/1538-4357/abc2d8} {\bibfield  {journal} {\bibinfo  {journal}
  {Astrophys. J.}\ }\textbf {\bibinfo {volume} {905}},\ \bibinfo {pages} {76}
  (\bibinfo {year} {2020})},\ \Eprint {http://arxiv.org/abs/2007.08582}
  {arXiv:2007.08582 [astro-ph.HE]} \BibitemShut {NoStop}%
\bibitem [{\citenamefont {Addazi}\ \emph {et~al.}(2022)\citenamefont {Addazi}
  \emph {et~al.}}]{LHAASO:2019qtb}%
  \BibitemOpen
  \bibfield  {author} {\bibinfo {author} {\bibfnamefont {A.}~\bibnamefont
  {Addazi}} \emph {et~al.} (\bibinfo {collaboration} {LHAASO}),\ }\href@noop {}
  {\bibfield  {journal} {\bibinfo  {journal} {Chin. Phys. C}\ }\textbf
  {\bibinfo {volume} {46}},\ \bibinfo {pages} {035001} (\bibinfo {year}
  {2022})},\ \Eprint {http://arxiv.org/abs/1905.02773} {arXiv:1905.02773
  [astro-ph.HE]} \BibitemShut {NoStop}%
\bibitem [{\citenamefont {{Cao}}\ \emph {et~al.}(2021)\citenamefont {{Cao}},
  \citenamefont {{Aharonian}}, \citenamefont {{An}}, \citenamefont {{Axikegu}},
  \citenamefont {{Bai}}, \citenamefont {{Bao}} \emph
  {et~al.}}]{2021Natur.594...33C}%
  \BibitemOpen
  \bibfield  {author} {\bibinfo {author} {\bibfnamefont {Z.}~\bibnamefont
  {{Cao}}}, \bibinfo {author} {\bibfnamefont {F.~A.}\ \bibnamefont
  {{Aharonian}}}, \bibinfo {author} {\bibfnamefont {Q.}~\bibnamefont {{An}}},
  \bibinfo {author} {\bibfnamefont {L.~X.}\ \bibnamefont {{Axikegu}},
  \bibfnamefont {Bai}}, \bibinfo {author} {\bibfnamefont {Y.~X.}\ \bibnamefont
  {{Bai}}}, \bibinfo {author} {\bibfnamefont {Y.~W.}\ \bibnamefont {{Bao}}},
  \emph {et~al.},\ }\href {\doibase 10.1038/s41586-021-03498-z} {\bibfield
  {journal} {\bibinfo  {journal} {\nat}\ }\textbf {\bibinfo {volume} {594}},\
  \bibinfo {pages} {33} (\bibinfo {year} {2021})}\BibitemShut {NoStop}%
\bibitem [{\citenamefont {Di~Mauro}\ \emph {et~al.}(2014)\citenamefont
  {Di~Mauro}, \citenamefont {Calore}, \citenamefont {Donato}, \citenamefont
  {Ajello},\ and\ \citenamefont {Latronico}}]{DiMauro:2013xta}%
  \BibitemOpen
  \bibfield  {author} {\bibinfo {author} {\bibfnamefont {M.}~\bibnamefont
  {Di~Mauro}}, \bibinfo {author} {\bibfnamefont {F.}~\bibnamefont {Calore}},
  \bibinfo {author} {\bibfnamefont {F.}~\bibnamefont {Donato}}, \bibinfo
  {author} {\bibfnamefont {M.}~\bibnamefont {Ajello}}, \ and\ \bibinfo {author}
  {\bibfnamefont {L.}~\bibnamefont {Latronico}},\ }\href {\doibase
  10.1088/0004-637X/780/2/161} {\bibfield  {journal} {\bibinfo  {journal}
  {Astrophys. J.}\ }\textbf {\bibinfo {volume} {780}},\ \bibinfo {pages} {161}
  (\bibinfo {year} {2014})},\ \Eprint {http://arxiv.org/abs/1304.0908}
  {arXiv:1304.0908 [astro-ph.HE]} \BibitemShut {NoStop}%
\bibitem [{\citenamefont {Tamborra}\ \emph {et~al.}(2014)\citenamefont
  {Tamborra}, \citenamefont {Ando},\ and\ \citenamefont
  {Murase}}]{Tamborra:2014xia}%
  \BibitemOpen
  \bibfield  {author} {\bibinfo {author} {\bibfnamefont {I.}~\bibnamefont
  {Tamborra}}, \bibinfo {author} {\bibfnamefont {S.}~\bibnamefont {Ando}}, \
  and\ \bibinfo {author} {\bibfnamefont {K.}~\bibnamefont {Murase}},\ }\href
  {\doibase 10.1088/1475-7516/2014/09/043} {\bibfield  {journal} {\bibinfo
  {journal} {JCAP}\ }\textbf {\bibinfo {volume} {09}},\ \bibinfo {pages} {043}
  (\bibinfo {year} {2014})},\ \Eprint {http://arxiv.org/abs/1404.1189}
  {arXiv:1404.1189 [astro-ph.HE]} \BibitemShut {NoStop}%
\bibitem [{\citenamefont {Roth}\ \emph {et~al.}(2021)\citenamefont {Roth},
  \citenamefont {Krumholz}, \citenamefont {Crocker},\ and\ \citenamefont
  {Celli}}]{Roth:2021lvk}%
  \BibitemOpen
  \bibfield  {author} {\bibinfo {author} {\bibfnamefont {M.~A.}\ \bibnamefont
  {Roth}}, \bibinfo {author} {\bibfnamefont {M.~R.}\ \bibnamefont {Krumholz}},
  \bibinfo {author} {\bibfnamefont {R.~M.}\ \bibnamefont {Crocker}}, \ and\
  \bibinfo {author} {\bibfnamefont {S.}~\bibnamefont {Celli}},\ }\href
  {\doibase 10.1038/s41586-021-03802-x} {\bibfield  {journal} {\bibinfo
  {journal} {Nature}\ }\textbf {\bibinfo {volume} {597}},\ \bibinfo {pages}
  {341} (\bibinfo {year} {2021})},\ \Eprint {http://arxiv.org/abs/2109.07598}
  {arXiv:2109.07598 [astro-ph.HE]} \BibitemShut {NoStop}%
\bibitem [{\citenamefont {Fornasa}\ and\ \citenamefont
  {S\'anchez-Conde}(2015)}]{Fornasa:2015qua}%
  \BibitemOpen
  \bibfield  {author} {\bibinfo {author} {\bibfnamefont {M.}~\bibnamefont
  {Fornasa}}\ and\ \bibinfo {author} {\bibfnamefont {M.~A.}\ \bibnamefont
  {S\'anchez-Conde}},\ }\href {\doibase 10.1016/j.physrep.2015.09.002}
  {\bibfield  {journal} {\bibinfo  {journal} {Phys. Rept.}\ }\textbf {\bibinfo
  {volume} {598}},\ \bibinfo {pages} {1} (\bibinfo {year} {2015})},\ \Eprint
  {http://arxiv.org/abs/1502.02866} {arXiv:1502.02866 [astro-ph.CO]}
  \BibitemShut {NoStop}%
\bibitem [{\citenamefont {Di~Mauro}\ and\ \citenamefont
  {Donato}(2015)}]{DiMauro:2015tfa}%
  \BibitemOpen
  \bibfield  {author} {\bibinfo {author} {\bibfnamefont {M.}~\bibnamefont
  {Di~Mauro}}\ and\ \bibinfo {author} {\bibfnamefont {F.}~\bibnamefont
  {Donato}},\ }\href {\doibase 10.1103/PhysRevD.91.123001} {\bibfield
  {journal} {\bibinfo  {journal} {Phys. Rev. D}\ }\textbf {\bibinfo {volume}
  {91}},\ \bibinfo {pages} {123001} (\bibinfo {year} {2015})},\ \Eprint
  {http://arxiv.org/abs/1501.05316} {arXiv:1501.05316 [astro-ph.HE]}
  \BibitemShut {NoStop}%
\bibitem [{\citenamefont {Ajello}\ \emph {et~al.}(2019)\citenamefont {Ajello}
  \emph {et~al.}}]{Ajello:2019zki}%
  \BibitemOpen
  \bibfield  {author} {\bibinfo {author} {\bibfnamefont {M.}~\bibnamefont
  {Ajello}} \emph {et~al.},\ }\href {\doibase 10.3847/1538-4357/ab1d4e}
  {\bibfield  {journal} {\bibinfo  {journal} {Astrophys. J.}\ }\textbf
  {\bibinfo {volume} {878}},\ \bibinfo {pages} {52} (\bibinfo {year} {2019})},\
  \Eprint {http://arxiv.org/abs/1906.11403} {arXiv:1906.11403 [astro-ph.HE]}
  \BibitemShut {NoStop}%
\bibitem [{\citenamefont {Ackermann}\ \emph {et~al.}(2012)\citenamefont
  {Ackermann} \emph {et~al.}}]{Fermi-LAT:2012edv}%
  \BibitemOpen
  \bibfield  {author} {\bibinfo {author} {\bibfnamefont {M.}~\bibnamefont
  {Ackermann}} \emph {et~al.} (\bibinfo {collaboration} {Fermi-LAT}),\ }\href
  {\doibase 10.1088/0004-637X/750/1/3} {\bibfield  {journal} {\bibinfo
  {journal} {Astrophys. J.}\ }\textbf {\bibinfo {volume} {750}},\ \bibinfo
  {pages} {3} (\bibinfo {year} {2012})},\ \Eprint
  {http://arxiv.org/abs/1202.4039} {arXiv:1202.4039 [astro-ph.HE]} \BibitemShut
  {NoStop}%
\bibitem [{\citenamefont {Acero}\ \emph {et~al.}(2016)\citenamefont {Acero}
  \emph {et~al.}}]{Fermi-LAT:2016zaq}%
  \BibitemOpen
  \bibfield  {author} {\bibinfo {author} {\bibfnamefont {F.}~\bibnamefont
  {Acero}} \emph {et~al.} (\bibinfo {collaboration} {Fermi-LAT}),\ }\href
  {\doibase 10.3847/0067-0049/223/2/26} {\bibfield  {journal} {\bibinfo
  {journal} {Astrophys. J. Suppl.}\ }\textbf {\bibinfo {volume} {223}},\
  \bibinfo {pages} {26} (\bibinfo {year} {2016})},\ \Eprint
  {http://arxiv.org/abs/1602.07246} {arXiv:1602.07246 [astro-ph.HE]}
  \BibitemShut {NoStop}%
\bibitem [{\citenamefont {Porter}\ \emph {et~al.}(2017)\citenamefont {Porter},
  \citenamefont {Johannesson},\ and\ \citenamefont
  {Moskalenko}}]{Porter:2017vaa}%
  \BibitemOpen
  \bibfield  {author} {\bibinfo {author} {\bibfnamefont {T.~A.}\ \bibnamefont
  {Porter}}, \bibinfo {author} {\bibfnamefont {G.}~\bibnamefont {Johannesson}},
  \ and\ \bibinfo {author} {\bibfnamefont {I.~V.}\ \bibnamefont {Moskalenko}},\
  }\href {\doibase 10.3847/1538-4357/aa844d} {\bibfield  {journal} {\bibinfo
  {journal} {Astrophys. J.}\ }\textbf {\bibinfo {volume} {846}},\ \bibinfo
  {pages} {67} (\bibinfo {year} {2017})},\ \Eprint
  {http://arxiv.org/abs/1708.00816} {arXiv:1708.00816 [astro-ph.HE]}
  \BibitemShut {NoStop}%
\bibitem [{\citenamefont {Kissmann}\ \emph {et~al.}(2017)\citenamefont
  {Kissmann}, \citenamefont {Niederwanger}, \citenamefont {Reimer},\ and\
  \citenamefont {Strong}}]{Kissmann:2017ghg}%
  \BibitemOpen
  \bibfield  {author} {\bibinfo {author} {\bibfnamefont {R.}~\bibnamefont
  {Kissmann}}, \bibinfo {author} {\bibfnamefont {F.}~\bibnamefont
  {Niederwanger}}, \bibinfo {author} {\bibfnamefont {O.}~\bibnamefont
  {Reimer}}, \ and\ \bibinfo {author} {\bibfnamefont {A.~W.}\ \bibnamefont
  {Strong}},\ }\href {\doibase 10.1063/1.4969008} {\bibfield  {journal}
  {\bibinfo  {journal} {AIP Conf. Proc.}\ }\textbf {\bibinfo {volume} {1792}},\
  \bibinfo {pages} {070011} (\bibinfo {year} {2017})},\ \Eprint
  {http://arxiv.org/abs/1701.07285} {arXiv:1701.07285 [astro-ph.HE]}
  \BibitemShut {NoStop}%
\bibitem [{\citenamefont {J\'ohannesson}\ \emph {et~al.}(2018)\citenamefont
  {J\'ohannesson}, \citenamefont {Porter},\ and\ \citenamefont
  {Moskalenko}}]{Johannesson:2018bit}%
  \BibitemOpen
  \bibfield  {author} {\bibinfo {author} {\bibfnamefont {G.}~\bibnamefont
  {J\'ohannesson}}, \bibinfo {author} {\bibfnamefont {T.~A.}\ \bibnamefont
  {Porter}}, \ and\ \bibinfo {author} {\bibfnamefont {I.~V.}\ \bibnamefont
  {Moskalenko}},\ }\href {\doibase 10.3847/1538-4357/aab26e} {\bibfield
  {journal} {\bibinfo  {journal} {Astrophys. J.}\ }\textbf {\bibinfo {volume}
  {856}},\ \bibinfo {pages} {45} (\bibinfo {year} {2018})},\ \Eprint
  {http://arxiv.org/abs/1802.08646} {arXiv:1802.08646 [astro-ph.HE]}
  \BibitemShut {NoStop}%
\bibitem [{\citenamefont {Tibaldo}\ \emph {et~al.}(2021)\citenamefont
  {Tibaldo}, \citenamefont {Gaggero},\ and\ \citenamefont
  {Martin}}]{Tibaldo:2021viq}%
  \BibitemOpen
  \bibfield  {author} {\bibinfo {author} {\bibfnamefont {L.}~\bibnamefont
  {Tibaldo}}, \bibinfo {author} {\bibfnamefont {D.}~\bibnamefont {Gaggero}}, \
  and\ \bibinfo {author} {\bibfnamefont {P.}~\bibnamefont {Martin}},\ }\href
  {\doibase 10.3390/universe7050141} {\bibfield  {journal} {\bibinfo  {journal}
  {Universe}\ }\textbf {\bibinfo {volume} {7}},\ \bibinfo {pages} {141}
  (\bibinfo {year} {2021})},\ \Eprint {http://arxiv.org/abs/2103.16423}
  {arXiv:2103.16423 [astro-ph.HE]} \BibitemShut {NoStop}%
\bibitem [{\citenamefont {Dundovic}\ \emph {et~al.}(2021)\citenamefont
  {Dundovic}, \citenamefont {Evoli}, \citenamefont {Gaggero},\ and\
  \citenamefont {Grasso}}]{Dundovic:2021ryb}%
  \BibitemOpen
  \bibfield  {author} {\bibinfo {author} {\bibfnamefont {A.}~\bibnamefont
  {Dundovic}}, \bibinfo {author} {\bibfnamefont {C.}~\bibnamefont {Evoli}},
  \bibinfo {author} {\bibfnamefont {D.}~\bibnamefont {Gaggero}}, \ and\
  \bibinfo {author} {\bibfnamefont {D.}~\bibnamefont {Grasso}},\ }\href
  {\doibase 10.1051/0004-6361/202140801} {\bibfield  {journal} {\bibinfo
  {journal} {Astron. Astrophys.}\ }\textbf {\bibinfo {volume} {653}},\ \bibinfo
  {pages} {A18} (\bibinfo {year} {2021})},\ \Eprint
  {http://arxiv.org/abs/2105.13165} {arXiv:2105.13165 [astro-ph.HE]}
  \BibitemShut {NoStop}%
\bibitem [{\citenamefont {Widmark}\ \emph {et~al.}(2022)\citenamefont
  {Widmark}, \citenamefont {Korsmeier},\ and\ \citenamefont
  {Linden}}]{Widmark:2022qgx}%
  \BibitemOpen
  \bibfield  {author} {\bibinfo {author} {\bibfnamefont {A.}~\bibnamefont
  {Widmark}}, \bibinfo {author} {\bibfnamefont {M.}~\bibnamefont {Korsmeier}},
  \ and\ \bibinfo {author} {\bibfnamefont {T.}~\bibnamefont {Linden}},\
  }\href@noop {} {\  (\bibinfo {year} {2022})},\ \Eprint
  {http://arxiv.org/abs/2208.11704} {arXiv:2208.11704 [astro-ph.GA]}
  \BibitemShut {NoStop}%
\bibitem [{\citenamefont {Strong}\ \emph {et~al.}(2000)\citenamefont {Strong},
  \citenamefont {Moskalenko},\ and\ \citenamefont {Reimer}}]{Strong:1998fr}%
  \BibitemOpen
  \bibfield  {author} {\bibinfo {author} {\bibfnamefont {A.~W.}\ \bibnamefont
  {Strong}}, \bibinfo {author} {\bibfnamefont {I.~V.}\ \bibnamefont
  {Moskalenko}}, \ and\ \bibinfo {author} {\bibfnamefont {O.}~\bibnamefont
  {Reimer}},\ }\href {\doibase 10.1086/309038} {\bibfield  {journal} {\bibinfo
  {journal} {Astrophys. J.}\ }\textbf {\bibinfo {volume} {537}},\ \bibinfo
  {pages} {763} (\bibinfo {year} {2000})},\ \bibinfo {note} {[Erratum:
  Astrophys.J. 541, 1109 (2000)]},\ \Eprint
  {http://arxiv.org/abs/astro-ph/9811296} {arXiv:astro-ph/9811296} \BibitemShut
  {NoStop}%
\bibitem [{\citenamefont {Aguilar}\ \emph {et~al.}(2021)\citenamefont {Aguilar}
  \emph {et~al.}}]{AMS:2021nhj}%
  \BibitemOpen
  \bibfield  {author} {\bibinfo {author} {\bibfnamefont {M.}~\bibnamefont
  {Aguilar}} \emph {et~al.} (\bibinfo {collaboration} {AMS}),\ }\href {\doibase
  10.1016/j.physrep.2020.09.003} {\bibfield  {journal} {\bibinfo  {journal}
  {Phys. Rept.}\ }\textbf {\bibinfo {volume} {894}},\ \bibinfo {pages} {1}
  (\bibinfo {year} {2021})}\BibitemShut {NoStop}%
\bibitem [{\citenamefont {Pohl}\ \emph {et~al.}(2008)\citenamefont {Pohl},
  \citenamefont {Englmaier},\ and\ \citenamefont {Bissantz}}]{Pohl:2007dz}%
  \BibitemOpen
  \bibfield  {author} {\bibinfo {author} {\bibfnamefont {M.}~\bibnamefont
  {Pohl}}, \bibinfo {author} {\bibfnamefont {P.}~\bibnamefont {Englmaier}}, \
  and\ \bibinfo {author} {\bibfnamefont {N.}~\bibnamefont {Bissantz}},\ }\href
  {\doibase 10.1086/529004} {\bibfield  {journal} {\bibinfo  {journal}
  {Astrophys. J.}\ }\textbf {\bibinfo {volume} {677}},\ \bibinfo {pages} {283}
  (\bibinfo {year} {2008})},\ \Eprint {http://arxiv.org/abs/0712.4264}
  {arXiv:0712.4264 [astro-ph]} \BibitemShut {NoStop}%
\bibitem [{\citenamefont {Mertsch}\ and\ \citenamefont
  {Phan}(2022)}]{Mertsch:2022oee}%
  \BibitemOpen
  \bibfield  {author} {\bibinfo {author} {\bibfnamefont {P.}~\bibnamefont
  {Mertsch}}\ and\ \bibinfo {author} {\bibfnamefont {V.~H.~M.}\ \bibnamefont
  {Phan}},\ }\href@noop {} {\  (\bibinfo {year} {2022})},\ \Eprint
  {http://arxiv.org/abs/2202.02341} {arXiv:2202.02341 [astro-ph.GA]}
  \BibitemShut {NoStop}%
\bibitem [{\citenamefont {Ackermann}\ \emph {et~al.}(2015)\citenamefont
  {Ackermann} \emph {et~al.}}]{Fermi-LAT:2014ryh}%
  \BibitemOpen
  \bibfield  {author} {\bibinfo {author} {\bibfnamefont {M.}~\bibnamefont
  {Ackermann}} \emph {et~al.} (\bibinfo {collaboration} {Fermi-LAT}),\ }\href
  {\doibase 10.1088/0004-637X/799/1/86} {\bibfield  {journal} {\bibinfo
  {journal} {Astrophys. J.}\ }\textbf {\bibinfo {volume} {799}},\ \bibinfo
  {pages} {86} (\bibinfo {year} {2015})},\ \Eprint
  {http://arxiv.org/abs/1410.3696} {arXiv:1410.3696 [astro-ph.HE]} \BibitemShut
  {NoStop}%
\bibitem [{\citenamefont {Yang}\ \emph {et~al.}(2016)\citenamefont {Yang},
  \citenamefont {Aharonian},\ and\ \citenamefont {Evoli}}]{Yang:2016jda}%
  \BibitemOpen
  \bibfield  {author} {\bibinfo {author} {\bibfnamefont {R.}~\bibnamefont
  {Yang}}, \bibinfo {author} {\bibfnamefont {F.}~\bibnamefont {Aharonian}}, \
  and\ \bibinfo {author} {\bibfnamefont {C.}~\bibnamefont {Evoli}},\ }\href
  {\doibase 10.1103/PhysRevD.93.123007} {\bibfield  {journal} {\bibinfo
  {journal} {Phys. Rev. D}\ }\textbf {\bibinfo {volume} {93}},\ \bibinfo
  {pages} {123007} (\bibinfo {year} {2016})},\ \Eprint
  {http://arxiv.org/abs/1602.04710} {arXiv:1602.04710 [astro-ph.HE]}
  \BibitemShut {NoStop}%
\bibitem [{\citenamefont {Pothast}\ \emph {et~al.}(2018)\citenamefont
  {Pothast}, \citenamefont {Gaggero}, \citenamefont {Storm},\ and\
  \citenamefont {Weniger}}]{Pothast:2018bvh}%
  \BibitemOpen
  \bibfield  {author} {\bibinfo {author} {\bibfnamefont {M.}~\bibnamefont
  {Pothast}}, \bibinfo {author} {\bibfnamefont {D.}~\bibnamefont {Gaggero}},
  \bibinfo {author} {\bibfnamefont {E.}~\bibnamefont {Storm}}, \ and\ \bibinfo
  {author} {\bibfnamefont {C.}~\bibnamefont {Weniger}},\ }\href {\doibase
  10.1088/1475-7516/2018/10/045} {\bibfield  {journal} {\bibinfo  {journal}
  {JCAP}\ }\textbf {\bibinfo {volume} {10}},\ \bibinfo {pages} {045} (\bibinfo
  {year} {2018})},\ \Eprint {http://arxiv.org/abs/1807.04554} {arXiv:1807.04554
  [astro-ph.HE]} \BibitemShut {NoStop}%
\bibitem [{\citenamefont {Kamae}\ \emph {et~al.}(2006)\citenamefont {Kamae},
  \citenamefont {Karlsson}, \citenamefont {Mizuno}, \citenamefont {Abe},\ and\
  \citenamefont {Koi}}]{Kamae:2006bf}%
  \BibitemOpen
  \bibfield  {author} {\bibinfo {author} {\bibfnamefont {T.}~\bibnamefont
  {Kamae}}, \bibinfo {author} {\bibfnamefont {N.}~\bibnamefont {Karlsson}},
  \bibinfo {author} {\bibfnamefont {T.}~\bibnamefont {Mizuno}}, \bibinfo
  {author} {\bibfnamefont {T.}~\bibnamefont {Abe}}, \ and\ \bibinfo {author}
  {\bibfnamefont {T.}~\bibnamefont {Koi}},\ }\href {\doibase 10.1086/513602}
  {\bibfield  {journal} {\bibinfo  {journal} {Astrophys. J.}\ }\textbf
  {\bibinfo {volume} {647}},\ \bibinfo {pages} {692} (\bibinfo {year}
  {2006})},\ \bibinfo {note} {[Erratum: Astrophys.J. 662, 779 (2007)]},\
  \Eprint {http://arxiv.org/abs/astro-ph/0605581} {arXiv:astro-ph/0605581}
  \BibitemShut {NoStop}%
\bibitem [{\citenamefont {Kachelrie\ss{}}\ \emph {et~al.}(2019)\citenamefont
  {Kachelrie\ss{}}, \citenamefont {Moskalenko},\ and\ \citenamefont
  {Ostapchenko}}]{Kachelriess:2019ifk}%
  \BibitemOpen
  \bibfield  {author} {\bibinfo {author} {\bibfnamefont {M.}~\bibnamefont
  {Kachelrie\ss{}}}, \bibinfo {author} {\bibfnamefont {I.~V.}\ \bibnamefont
  {Moskalenko}}, \ and\ \bibinfo {author} {\bibfnamefont {S.}~\bibnamefont
  {Ostapchenko}},\ }\href {\doibase 10.1016/j.cpc.2019.08.001} {\bibfield
  {journal} {\bibinfo  {journal} {Comput. Phys. Commun.}\ }\textbf {\bibinfo
  {volume} {245}},\ \bibinfo {pages} {106846} (\bibinfo {year} {2019})},\
  \Eprint {http://arxiv.org/abs/1904.05129} {arXiv:1904.05129 [hep-ph]}
  \BibitemShut {NoStop}%
\bibitem [{\citenamefont {Bhatt}\ \emph {et~al.}(2020)\citenamefont {Bhatt},
  \citenamefont {Sushch}, \citenamefont {Pohl}, \citenamefont {Fedynitch},
  \citenamefont {Das}, \citenamefont {Brose}, \citenamefont {Plotko},\ and\
  \citenamefont {Meyer}}]{Bhatt_2020}%
  \BibitemOpen
  \bibfield  {author} {\bibinfo {author} {\bibfnamefont {M.}~\bibnamefont
  {Bhatt}}, \bibinfo {author} {\bibfnamefont {I.}~\bibnamefont {Sushch}},
  \bibinfo {author} {\bibfnamefont {M.}~\bibnamefont {Pohl}}, \bibinfo {author}
  {\bibfnamefont {A.}~\bibnamefont {Fedynitch}}, \bibinfo {author}
  {\bibfnamefont {S.}~\bibnamefont {Das}}, \bibinfo {author} {\bibfnamefont
  {R.}~\bibnamefont {Brose}}, \bibinfo {author} {\bibfnamefont
  {P.}~\bibnamefont {Plotko}}, \ and\ \bibinfo {author} {\bibfnamefont
  {D.~M.-A.}\ \bibnamefont {Meyer}},\ }\href {\doibase
  10.1016/j.astropartphys.2020.102490} {\bibfield  {journal} {\bibinfo
  {journal} {Astroparticle Physics}\ }\textbf {\bibinfo {volume} {123}},\
  \bibinfo {pages} {102490} (\bibinfo {year} {2020})}\BibitemShut {NoStop}%
\bibitem [{\citenamefont {Mazziotta}\ \emph {et~al.}(2016)\citenamefont
  {Mazziotta}, \citenamefont {Cerutti}, \citenamefont {Ferrari}, \citenamefont
  {Gaggero}, \citenamefont {Loparco},\ and\ \citenamefont
  {Sala}}]{Mazziotta_2016}%
  \BibitemOpen
  \bibfield  {author} {\bibinfo {author} {\bibfnamefont {M.}~\bibnamefont
  {Mazziotta}}, \bibinfo {author} {\bibfnamefont {F.}~\bibnamefont {Cerutti}},
  \bibinfo {author} {\bibfnamefont {A.}~\bibnamefont {Ferrari}}, \bibinfo
  {author} {\bibfnamefont {D.}~\bibnamefont {Gaggero}}, \bibinfo {author}
  {\bibfnamefont {F.}~\bibnamefont {Loparco}}, \ and\ \bibinfo {author}
  {\bibfnamefont {P.}~\bibnamefont {Sala}},\ }\href {\doibase
  10.1016/j.astropartphys.2016.04.005} {\bibfield  {journal} {\bibinfo
  {journal} {Astroparticle Physics}\ }\textbf {\bibinfo {volume} {81}},\
  \bibinfo {pages} {21} (\bibinfo {year} {2016})}\BibitemShut {NoStop}%
\bibitem [{\citenamefont {Kachelriess}\ \emph {et~al.}(2015)\citenamefont
  {Kachelriess}, \citenamefont {Moskalenko},\ and\ \citenamefont
  {Ostapchenko}}]{Kachelriess:2015wpa}%
  \BibitemOpen
  \bibfield  {author} {\bibinfo {author} {\bibfnamefont {M.}~\bibnamefont
  {Kachelriess}}, \bibinfo {author} {\bibfnamefont {I.~V.}\ \bibnamefont
  {Moskalenko}}, \ and\ \bibinfo {author} {\bibfnamefont {S.~S.}\ \bibnamefont
  {Ostapchenko}},\ }\href {\doibase 10.1088/0004-637X/803/2/54} {\bibfield
  {journal} {\bibinfo  {journal} {Astrophys. J.}\ }\textbf {\bibinfo {volume}
  {803}},\ \bibinfo {pages} {54} (\bibinfo {year} {2015})},\ \Eprint
  {http://arxiv.org/abs/1502.04158} {arXiv:1502.04158 [astro-ph.HE]}
  \BibitemShut {NoStop}%
\bibitem [{\citenamefont {Kachelrie\ss{}}\ \emph {et~al.}(2020)\citenamefont
  {Kachelrie\ss{}}, \citenamefont {Ostapchenko},\ and\ \citenamefont
  {Tjemsland}}]{Kachelriess:2019taq}%
  \BibitemOpen
  \bibfield  {author} {\bibinfo {author} {\bibfnamefont {M.}~\bibnamefont
  {Kachelrie\ss{}}}, \bibinfo {author} {\bibfnamefont {S.}~\bibnamefont
  {Ostapchenko}}, \ and\ \bibinfo {author} {\bibfnamefont {J.}~\bibnamefont
  {Tjemsland}},\ }\href {\doibase 10.1140/epja/s10050-019-00007-9} {\bibfield
  {journal} {\bibinfo  {journal} {Eur. Phys. J. A}\ }\textbf {\bibinfo {volume}
  {56}},\ \bibinfo {pages} {4} (\bibinfo {year} {2020})},\ \Eprint
  {http://arxiv.org/abs/1905.01192} {arXiv:1905.01192 [hep-ph]} \BibitemShut
  {NoStop}%
\bibitem [{\citenamefont {Orusa}\ \emph {et~al.}(2022)\citenamefont {Orusa},
  \citenamefont {Di~Mauro}, \citenamefont {Donato},\ and\ \citenamefont
  {Korsmeier}}]{Orusa:2022pvp}%
  \BibitemOpen
  \bibfield  {author} {\bibinfo {author} {\bibfnamefont {L.}~\bibnamefont
  {Orusa}}, \bibinfo {author} {\bibfnamefont {M.}~\bibnamefont {Di~Mauro}},
  \bibinfo {author} {\bibfnamefont {F.}~\bibnamefont {Donato}}, \ and\ \bibinfo
  {author} {\bibfnamefont {M.}~\bibnamefont {Korsmeier}},\ }\href {\doibase
  10.1103/PhysRevD.105.123021} {\bibfield  {journal} {\bibinfo  {journal}
  {Phys. Rev. D}\ }\textbf {\bibinfo {volume} {105}},\ \bibinfo {pages}
  {123021} (\bibinfo {year} {2022})},\ \Eprint
  {http://arxiv.org/abs/2203.13143} {arXiv:2203.13143 [astro-ph.HE]}
  \BibitemShut {NoStop}%
\bibitem [{\citenamefont {Koldobskiy}\ \emph {et~al.}(2021)\citenamefont
  {Koldobskiy}, \citenamefont {Kachelrie\ss{}}, \citenamefont {Lskavyan},
  \citenamefont {Neronov}, \citenamefont {Ostapchenko},\ and\ \citenamefont
  {Semikoz}}]{Koldobskiy:2021nld}%
  \BibitemOpen
  \bibfield  {author} {\bibinfo {author} {\bibfnamefont {S.}~\bibnamefont
  {Koldobskiy}}, \bibinfo {author} {\bibfnamefont {M.}~\bibnamefont
  {Kachelrie\ss{}}}, \bibinfo {author} {\bibfnamefont {A.}~\bibnamefont
  {Lskavyan}}, \bibinfo {author} {\bibfnamefont {A.}~\bibnamefont {Neronov}},
  \bibinfo {author} {\bibfnamefont {S.}~\bibnamefont {Ostapchenko}}, \ and\
  \bibinfo {author} {\bibfnamefont {D.~V.}\ \bibnamefont {Semikoz}},\ }\href
  {\doibase 10.1103/PhysRevD.104.123027} {\bibfield  {journal} {\bibinfo
  {journal} {Phys. Rev. D}\ }\textbf {\bibinfo {volume} {104}},\ \bibinfo
  {pages} {123027} (\bibinfo {year} {2021})},\ \Eprint
  {http://arxiv.org/abs/2110.00496} {arXiv:2110.00496 [astro-ph.HE]}
  \BibitemShut {NoStop}%
\bibitem [{\citenamefont {Moskalenko}\ and\ \citenamefont
  {Strong}(1998)}]{Moskalenko:1997gh}%
  \BibitemOpen
  \bibfield  {author} {\bibinfo {author} {\bibfnamefont {I.~V.}\ \bibnamefont
  {Moskalenko}}\ and\ \bibinfo {author} {\bibfnamefont {A.~W.}\ \bibnamefont
  {Strong}},\ }\href {\doibase 10.1086/305152} {\bibfield  {journal} {\bibinfo
  {journal} {Astrophys. J.}\ }\textbf {\bibinfo {volume} {493}},\ \bibinfo
  {pages} {694} (\bibinfo {year} {1998})},\ \Eprint
  {http://arxiv.org/abs/astro-ph/9710124} {arXiv:astro-ph/9710124} \BibitemShut
  {NoStop}%
\bibitem [{\citenamefont {Matthews}\ \emph {et~al.}(2020)\citenamefont
  {Matthews}, \citenamefont {Bell},\ and\ \citenamefont
  {Blundell}}]{Matthews:2020lig}%
  \BibitemOpen
  \bibfield  {author} {\bibinfo {author} {\bibfnamefont {J.}~\bibnamefont
  {Matthews}}, \bibinfo {author} {\bibfnamefont {A.}~\bibnamefont {Bell}}, \
  and\ \bibinfo {author} {\bibfnamefont {K.}~\bibnamefont {Blundell}},\ }\href
  {\doibase 10.1016/j.newar.2020.101543} {\bibfield  {journal} {\bibinfo
  {journal} {New Astron. Rev.}\ }\textbf {\bibinfo {volume} {89}},\ \bibinfo
  {pages} {101543} (\bibinfo {year} {2020})},\ \Eprint
  {http://arxiv.org/abs/2003.06587} {arXiv:2003.06587 [astro-ph.HE]}
  \BibitemShut {NoStop}%
\bibitem [{\citenamefont {Hooper}\ and\ \citenamefont
  {Goodenough}(2011)}]{Hooper:2010mq}%
  \BibitemOpen
  \bibfield  {author} {\bibinfo {author} {\bibfnamefont {D.}~\bibnamefont
  {Hooper}}\ and\ \bibinfo {author} {\bibfnamefont {L.}~\bibnamefont
  {Goodenough}},\ }\href {\doibase 10.1016/j.physletb.2011.02.029} {\bibfield
  {journal} {\bibinfo  {journal} {Phys. Lett. B}\ }\textbf {\bibinfo {volume}
  {697}},\ \bibinfo {pages} {412} (\bibinfo {year} {2011})},\ \Eprint
  {http://arxiv.org/abs/1010.2752} {arXiv:1010.2752 [hep-ph]} \BibitemShut
  {NoStop}%
\bibitem [{\citenamefont {Ackermann}\ \emph {et~al.}(2017)\citenamefont
  {Ackermann} \emph {et~al.}}]{Fermi-LAT:2017opo}%
  \BibitemOpen
  \bibfield  {author} {\bibinfo {author} {\bibfnamefont {M.}~\bibnamefont
  {Ackermann}} \emph {et~al.} (\bibinfo {collaboration} {Fermi-LAT}),\ }\href
  {\doibase 10.3847/1538-4357/aa6cab} {\bibfield  {journal} {\bibinfo
  {journal} {Astrophys. J.}\ }\textbf {\bibinfo {volume} {840}},\ \bibinfo
  {pages} {43} (\bibinfo {year} {2017})},\ \Eprint
  {http://arxiv.org/abs/1704.03910} {arXiv:1704.03910 [astro-ph.HE]}
  \BibitemShut {NoStop}%
\bibitem [{\citenamefont {Alt}\ \emph {et~al.}(2005)\citenamefont {Alt} \emph
  {et~al.}}]{2005_NA49}%
  \BibitemOpen
  \bibfield  {author} {\bibinfo {author} {\bibfnamefont {C.}~\bibnamefont
  {Alt}} \emph {et~al.} (\bibinfo {collaboration} {NA49}),\ }\href {\doibase
  10.1140/epjc/s2005-02391-9} {\bibfield  {journal} {\bibinfo  {journal} {Eur.
  Phys. J. C}\ }\textbf {\bibinfo {volume} {45}},\ \bibinfo {pages} {343–381}
  (\bibinfo {year} {2005})}\BibitemShut {NoStop}%
\bibitem [{\citenamefont {Aduszkiewicz}\ \emph {et~al.}(2017)\citenamefont
  {Aduszkiewicz} \emph {et~al.}}]{Aduszkiewicz:2017sei}%
  \BibitemOpen
  \bibfield  {author} {\bibinfo {author} {\bibfnamefont {A.}~\bibnamefont
  {Aduszkiewicz}} \emph {et~al.} (\bibinfo {collaboration} {NA61/SHINE}),\
  }\href {\doibase 10.1140/epjc/s10052-017-5260-4} {\bibfield  {journal}
  {\bibinfo  {journal} {Eur. Phys. J. C}\ }\textbf {\bibinfo {volume} {77}},\
  \bibinfo {pages} {671} (\bibinfo {year} {2017})},\ \Eprint
  {http://arxiv.org/abs/1705.02467} {arXiv:1705.02467 [nucl-ex]} \BibitemShut
  {NoStop}%
\bibitem [{\citenamefont {Dermer}(1986)}]{dermer1986binary}%
  \BibitemOpen
  \bibfield  {author} {\bibinfo {author} {\bibfnamefont {C.~D.}\ \bibnamefont
  {Dermer}},\ }\href@noop {} {\bibfield  {journal} {\bibinfo  {journal} {The
  Astrophysical Journal}\ }\textbf {\bibinfo {volume} {307}},\ \bibinfo {pages}
  {47} (\bibinfo {year} {1986})}\BibitemShut {NoStop}%
\bibitem [{\citenamefont {{Stecker}}(1973)}]{Stecker73}%
  \BibitemOpen
  \bibfield  {author} {\bibinfo {author} {\bibfnamefont {F.~W.}\ \bibnamefont
  {{Stecker}}},\ }\href {\doibase 10.1086/152435} {\bibfield  {journal}
  {\bibinfo  {journal} {\apj}\ }\textbf {\bibinfo {volume} {185}},\ \bibinfo
  {pages} {499} (\bibinfo {year} {1973})}\BibitemShut {NoStop}%
\bibitem [{\citenamefont {Adriani}\ \emph {et~al.}(2016)\citenamefont {Adriani}
  \emph {et~al.}}]{Adriani_2016}%
  \BibitemOpen
  \bibfield  {author} {\bibinfo {author} {\bibfnamefont {O.}~\bibnamefont
  {Adriani}} \emph {et~al.},\ }\href {\doibase 10.1103/physrevd.94.032007}
  {\bibfield  {journal} {\bibinfo  {journal} {Physical Review D}\ }\textbf
  {\bibinfo {volume} {94}} (\bibinfo {year} {2016}),\
  10.1103/physrevd.94.032007}\BibitemShut {NoStop}%
\bibitem [{\citenamefont {Acharya}\ \emph {et~al.}(2017)\citenamefont {Acharya}
  \emph {et~al.}}]{ALICE:2017nce}%
  \BibitemOpen
  \bibfield  {author} {\bibinfo {author} {\bibfnamefont {S.}~\bibnamefont
  {Acharya}} \emph {et~al.} (\bibinfo {collaboration} {ALICE}),\ }\href
  {\doibase 10.1140/epjc/s10052-017-4890-x} {\bibfield  {journal} {\bibinfo
  {journal} {Eur. Phys. J. C}\ }\textbf {\bibinfo {volume} {77}},\ \bibinfo
  {pages} {339} (\bibinfo {year} {2017})},\ \Eprint
  {http://arxiv.org/abs/1702.00917} {arXiv:1702.00917 [hep-ex]} \BibitemShut
  {NoStop}%
\bibitem [{\citenamefont {Feroz}\ \emph {et~al.}(2009)\citenamefont {Feroz},
  \citenamefont {Hobson},\ and\ \citenamefont {Bridges}}]{Multinest_2009}%
  \BibitemOpen
  \bibfield  {author} {\bibinfo {author} {\bibfnamefont {F.}~\bibnamefont
  {Feroz}}, \bibinfo {author} {\bibfnamefont {M.~P.}\ \bibnamefont {Hobson}}, \
  and\ \bibinfo {author} {\bibfnamefont {M.}~\bibnamefont {Bridges}},\ }\href
  {\doibase 10.1111/j.1365-2966.2009.14548.x} {\bibfield  {journal} {\bibinfo
  {journal} {Monthly Notices of the Royal Astronomical Society}\ }\textbf
  {\bibinfo {volume} {398}},\ \bibinfo {pages} {1601–1614} (\bibinfo {year}
  {2009})}\BibitemShut {NoStop}%
\bibitem [{\citenamefont {Korsmeier}\ \emph {et~al.}(2018)\citenamefont
  {Korsmeier}, \citenamefont {Donato},\ and\ \citenamefont
  {Di~Mauro}}]{Korsmeier_2018}%
  \BibitemOpen
  \bibfield  {author} {\bibinfo {author} {\bibfnamefont {M.}~\bibnamefont
  {Korsmeier}}, \bibinfo {author} {\bibfnamefont {F.}~\bibnamefont {Donato}}, \
  and\ \bibinfo {author} {\bibfnamefont {M.}~\bibnamefont {Di~Mauro}},\ }\href
  {\doibase 10.1103/physrevd.97.103019} {\bibfield  {journal} {\bibinfo
  {journal} {Physical Review D}\ }\textbf {\bibinfo {volume} {97}} (\bibinfo
  {year} {2018}),\ 10.1103/physrevd.97.103019}\BibitemShut {NoStop}%
\bibitem [{\citenamefont {Sj\"ostrand}\ \emph {et~al.}(2015)\citenamefont
  {Sj\"ostrand}, \citenamefont {Ask}, \citenamefont {Christiansen},
  \citenamefont {Corke}, \citenamefont {Desai}, \citenamefont {Ilten},
  \citenamefont {Mrenna}, \citenamefont {Prestel}, \citenamefont {Rasmussen},\
  and\ \citenamefont {Skands}}]{Sjostrand:2014zea}%
  \BibitemOpen
  \bibfield  {author} {\bibinfo {author} {\bibfnamefont {T.}~\bibnamefont
  {Sj\"ostrand}}, \bibinfo {author} {\bibfnamefont {S.}~\bibnamefont {Ask}},
  \bibinfo {author} {\bibfnamefont {J.~R.}\ \bibnamefont {Christiansen}},
  \bibinfo {author} {\bibfnamefont {R.}~\bibnamefont {Corke}}, \bibinfo
  {author} {\bibfnamefont {N.}~\bibnamefont {Desai}}, \bibinfo {author}
  {\bibfnamefont {P.}~\bibnamefont {Ilten}}, \bibinfo {author} {\bibfnamefont
  {S.}~\bibnamefont {Mrenna}}, \bibinfo {author} {\bibfnamefont
  {S.}~\bibnamefont {Prestel}}, \bibinfo {author} {\bibfnamefont {C.~O.}\
  \bibnamefont {Rasmussen}}, \ and\ \bibinfo {author} {\bibfnamefont {P.~Z.}\
  \bibnamefont {Skands}},\ }\href {\doibase 10.1016/j.cpc.2015.01.024}
  {\bibfield  {journal} {\bibinfo  {journal} {Comput. Phys. Commun.}\ }\textbf
  {\bibinfo {volume} {191}},\ \bibinfo {pages} {159} (\bibinfo {year}
  {2015})},\ \Eprint {http://arxiv.org/abs/1410.3012} {arXiv:1410.3012
  [hep-ph]} \BibitemShut {NoStop}%
\bibitem [{\citenamefont {Acharya}\ \emph {et~al.}(2018)\citenamefont {Acharya}
  \emph {et~al.}}]{ALICE:2017ryd}%
  \BibitemOpen
  \bibfield  {author} {\bibinfo {author} {\bibfnamefont {S.}~\bibnamefont
  {Acharya}} \emph {et~al.} (\bibinfo {collaboration} {ALICE}),\ }\href
  {\doibase 10.1140/epjc/s10052-018-5612-8} {\bibfield  {journal} {\bibinfo
  {journal} {Eur. Phys. J. C}\ }\textbf {\bibinfo {volume} {78}},\ \bibinfo
  {pages} {263} (\bibinfo {year} {2018})},\ \Eprint
  {http://arxiv.org/abs/1708.08745} {arXiv:1708.08745 [hep-ex]} \BibitemShut
  {NoStop}%
\bibitem [{\citenamefont {Adare}\ \emph {et~al.}(2011)\citenamefont {Adare}
  \emph {et~al.}}]{PHENIX:2010hvs}%
  \BibitemOpen
  \bibfield  {author} {\bibinfo {author} {\bibfnamefont {A.}~\bibnamefont
  {Adare}} \emph {et~al.} (\bibinfo {collaboration} {PHENIX}),\ }\href
  {\doibase 10.1103/PhysRevD.83.032001} {\bibfield  {journal} {\bibinfo
  {journal} {Phys. Rev. D}\ }\textbf {\bibinfo {volume} {83}},\ \bibinfo
  {pages} {032001} (\bibinfo {year} {2011})},\ \Eprint
  {http://arxiv.org/abs/1009.6224} {arXiv:1009.6224 [hep-ex]} \BibitemShut
  {NoStop}%
\bibitem [{\citenamefont {Albrecht}\ \emph {et~al.}(1995)\citenamefont
  {Albrecht} \emph {et~al.}}]{WA80:1995whm}%
  \BibitemOpen
  \bibfield  {author} {\bibinfo {author} {\bibfnamefont {R.}~\bibnamefont
  {Albrecht}} \emph {et~al.} (\bibinfo {collaboration} {WA80}),\ }\href
  {\doibase 10.1016/0370-2693(95)01166-N} {\bibfield  {journal} {\bibinfo
  {journal} {Phys. Lett. B}\ }\textbf {\bibinfo {volume} {361}},\ \bibinfo
  {pages} {14} (\bibinfo {year} {1995})},\ \Eprint
  {http://arxiv.org/abs/hep-ex/9507009} {arXiv:hep-ex/9507009} \BibitemShut
  {NoStop}%
\end{thebibliography}%

\end{document}